\documentclass[12pt]{article}

\usepackage[utf8]{inputenc}
\usepackage[english]{babel}
\usepackage{amsmath,amssymb,amsfonts,amsthm,hyperref}
\usepackage{graphicx,multirow,caption,bbm,latexsym,epsfig}
\usepackage{cite}
\usepackage{xcolor,colortbl}

\newcommand{\bs}{\begin{subequations}}
\newcommand{\es}{\end{subequations}}
\newcommand{\be}{\begin{equation}}
\newcommand{\ee}{\end{equation}}
\newcommand{\ba}{\begin{eqnarray}}
\newcommand{\ea}{\end{eqnarray}}
\newcommand{\no}{\nonumber \\}

\newcommand{\viz}{\textit{viz.}}
\newcommand{\ie}{\textit{i.e.}}

\def\no{\nonumber\\}

\allowdisplaybreaks

\graphicspath{ {figures/} }

\begin{document}

\title{\LARGE The oblique parameters from arbitrary new fermions}

\author{
  F.~Albergaria,$^{(1)}$\thanks{E-mail:
    \tt francisco.albergaria@tecnico.ulisboa.pt}
  \
  D.~Jur\v{c}iukonis,$^{(2)}$\thanks{E-mail:
    \tt darius.jurciukonis@tfai.vu.lt}
  \
  and L.~Lavoura$^{(1)}$\thanks{E-mail:
    \tt balio@cftp.tecnico.ulisboa.pt}
  \\*[3mm]
  $^{(1)}\!$
  \small Universidade de Lisboa, Instituto Superior T\'ecnico, CFTP, \\
  \small Av.~Rovisco~Pais~1, 1049-001~Lisboa, Portugal
  \\*[2mm]
  $^{(2)}\!$
  \small Vilnius University, Institute of Theoretical Physics and Astronomy, \\
  \small Saul\.etekio~av.~3, Vilnius 10257, Lithuania
}

\maketitle

\begin{abstract}
  We compute the six oblique parameters $S, T, U, V, W, X$ 
  in a New Physics Model with an arbitrary number of new fermions,
  in arbitrary representations of $SU(2) \times U(1)$,
  and mixing arbitrarily among themselves.
  We show that $S$ and $U$ are automatically finite,
  but $T$ is finite only if there is a specific relation
  between the masses of the new fermions
  and the representations of $SU(2) \times U(1)$ that they sit in.
  We apply our general computation to two illustrative cases.
  %  of an NPM with fermions in an arrangement
  %  of two vector-like representations of $SU(2) \times U(1)$
  %  such that the vacuum expectation value
  %  of the Standard Model's scalar doublet
  %  produces a mixing between them.
\end{abstract}

\section{Introduction}
\label{sec:introduction}

The oblique parameters (OPs) provide a convenient way of comparing
the predictions of a New Physics Model (NPM)
with those of the Standard Model (SM).
The NPM is supposed to have the same gauge group as the SM,
\viz~$SU(2) \times U(1)$.
The different particle content
between the NPM and the SM must consist solely
of extra fermions and/or scalars in the NPM.
Those new fermions and scalars
should preferably be in representations of the gauge group
such that they cannot couple to the light fermions
with which most experiments are performed;
in that way,
one ensures that their only effects are through their contributions
to the vacuum polarizations,
\ie~to the self-energies of the gauge bosons.
One writes those new contributions,
coming from loops\footnote{We only consider
the \emph{one}-loop level vacuum polarizations.}
of the extra fermions and/or scalars,
as
\be
\Pi^{\mu \nu}_{V V^\prime} \left( q \right)
= g^{\mu \nu} A_{V V^\prime} \left( q^2 \right)
+ q^\mu q^\nu\, B_{V V^\prime} \left( q^2 \right),
\ee
where $q^\mu$ is the four-momentum of the gauge bosons
and $V$ and $V^\prime$ are the gauge bosons at hand,
which may be either $W^+$ and $W^-$,
or a photon $\gamma$ and a $Z^0$,
or two photons,
or two $Z^0$'s.
%%%%% I ADDED:
Note that the functions $A_{V V^\prime} \left( q^2 \right)$
have mass-squared dimensions.
%%%%%
Let us denote
\bs
\ba
A^\prime_{V V^\prime} \left( q^2 \right)
&=&
\frac{\mathrm{d} A_{V V^\prime} \left( q^2 \right)}{\mathrm{d} q^2},
\\
\widetilde A_{V V^\prime} \left( q^2 \right)
&=&
\frac{A_{V V^\prime} \left( q^2 \right) - A_{V V^\prime} \left( 0 \right)}{q^2}.
\ea
\es
Then the OPs are defined as\footnote{In Eqs.~\eqref{fj9493}
we have used the sign conventions for $s_W$ and $c_W$
in Ref.~\cite{book}.
However,
the formulas that we shall present for the OPs
do not depend on those conventions.}$^,$\footnote{We adopt
the definitions of the OPs in Ref.~\cite{maksymyk}.
Those definitions do not neglect the second derivatives
of the $A_{V V^\prime} \left( q^2 \right)$ relative to $q^2$.
For this reason,
they produce extra parameters $V$, $W$, and $X$.}
\bs
\label{fj9493}
\ba
S &=& \frac{4 s_W^2 c_W^2}{\alpha}
\left[ \widetilde A_{ZZ} \left( m_Z^2 \right)
  + \frac{c_W^2 - s_W^2}{c_W s_W}\, A^\prime_{\gamma Z} \left( 0 \right)
  - A^\prime_{\gamma \gamma} \left( 0 \right) \right],
\label{sdef} \\
T &=& \frac{1}{\alpha} \left[
  \frac{A_{WW} \left( 0 \right)}{m_W^2}
  - \frac{A_{ZZ} \left( 0 \right)}{m_Z^2} \right],
\label{tdef} \\
U &=& - S + \frac{4 s_W^2}{\alpha} \left[ \widetilde A_{WW} \left( m_W^2 \right)
  + \frac{c_W}{s_W}\, A^\prime_{\gamma Z} \left( 0 \right)
  - A^\prime_{\gamma \gamma} \left( 0 \right) \right],
\label{udef} \\
V &=& \frac{1}{\alpha} \left[ A^\prime_{ZZ} \left( m_Z^2 \right)
  - \widetilde A_{ZZ} \left( m_Z^2 \right) \right],
\\
W &=& \frac{1}{\alpha} \left[ A^\prime_{WW} \left( m_W^2 \right)
  - \widetilde A_{WW} \left( m_W^2 \right) \right],
\\
X &=& \frac{s_W c_W}{\alpha} \left[ A^\prime_{\gamma Z} \left( 0 \right)
  - \widetilde A_{\gamma Z} \left( m_Z^2 \right) \right].
\ea
\es
In Eqs.~\eqref{fj9493},
$\alpha$ is the fine-structure constant,
$s_W$ and $c_W$ are the sine and the cosine,
respectively,
of the Weinberg angle $\theta_W$,
and $m_Z$ and $m_W$ are the masses of the $Z^0$ and $W^\pm$,
respectively.
At tree level
\be
\label{ibvugpgf}
m_W = c_W m_Z
\ee
in both the NPM and the SM;
this is because no neutral-scalar field
is allowed to acquire a vacuum expectation value (VEV)
unless it has either $J = Y = 0$ or $J = Y = 1/2$,\footnote{A few
other exceptional values of $J$ and $Y$,
like $J=3$ and $Y=2$,
are permitted too.}
where $J$ is the (total) weak isospin and $Y$
is the weak hypercharge.

The comparison between the predictions
of an NPM and the ones of the SM is done through formulas like,
for instance,
\be
\label{jvifof}
\frac{m_W^\mathrm{NPM}}{m_W^\mathrm{SM}} = 1 + \alpha \left[
  \frac{S}{4 \left( s_W^2 - c_W^S \right)}
  + \frac{c_W^2 T}{2 \left( c_W^2 - s_W^2 \right)}
  + \frac{U}{8 s_W^2} \right],
\ee
wherein the input observables
in the renormalization of both the SM and the NPM
are assumed to be $\alpha$,
$m_Z$,
and the Fermi coupling constant $G_F$
measured in muon decay;\footnote{The angle $\theta_W$
is extracted from these input observables through
\be
s_W^2 + c_W^2 = 1
\quad \mathrm{and} \quad
s_W c_W = \frac{\pi \alpha}{\sqrt{2} G_F m_Z^2}.
\ee
Equation~\eqref{ibvugpgf} is \emph{not} supposed to hold at loop level.}
the mass $m_W$ is thought of as a \emph{prediction}
of either the SM or the NPM.
Formulas analogous to Eq.~\eqref{jvifof} exist
for some twenty other measured observables~\cite{Drauksas}.

General formulas for the OPs
when the new fermions of the NPM are placed in either singlets,
doublets,
or triplets of $SU(2)$,
and have some specific hypercharges,
have been recently derived in Ref.~\cite{romao}.
General formulas for the OPs
when the new particles of the NPM are scalars in \emph{any} representations
of $SU(2) \times U(1)$ have been presented in Ref.~\cite{albergaria}.
Here we generalize both papers
by presenting general formulas for the OPs
when the new particles of the NPM are fermions
in \emph{any} representations
of $SU(2) \times U(1)$.
We allow the new fermions to have arbitrary masses
and to mix freely among themselves.\footnote{We do not consider mixing
between the NP fermions and the SM fermions.
If this mixing is present,
then one must do the computations of the OPs
by following the recipe we give here
both for the NPM and for the SM,
and,
afterwards,
the true OPs are given by
$\mathrm{OP} = \mathrm{OP}^\mathrm{NPM} - \mathrm{OP}^\mathrm{SM}$.}
We do not specify the mechanism through which the fermion masses are generated.
We implicitly assume the new fermions to be
of Dirac type.\footnote{Various interesting sets of
fermions that may be added to the SM have been identified
in Ref.~\cite{bizot}. Many of those sets contain
Majorana neutrinos.}

This paper is organized as follows.
In section~\ref{sec:functions} we introduce the functions
in terms of which we are later going to write down the OPs.
In section~\ref{sec:mixing} we define the mixing matrices of the fermions
and we prove some equations that apply to them.
The formulas for the oblique parameters are displayed in section~\ref{sec:ops};
we also demonstrate there the cancellation of the divergences of $S$ and $U$,
and we write down the equation that must be satisfied
in order for the divergence of $T$ to vanish too. 
In section~\ref{sec:one} we consider the simple case
of one vector-like multiplet of fermions,
while in section~\ref{sec:two} we analyse a model with
two vector-like multiplets of fermions.
We draw our conclusions in section~\ref{sec:conclusions}.
In Appendix~\ref{appa} we give formulas for the parameters $S$ and $U$
as they were defined in the original work by Peskin and Takeuchi~\cite{PT}.

\section{Functions}
\label{sec:functions}

In Ref.~\cite{romao} a few functions have been found
to be relevant to the formulas for the OPs
in an NPM with singlet,
doublet,
and triplet fermions.
Now we have found that those functions are,
indeed,
all that one needs to write down the OPs
when there are \emph{any} new fermions.
The functions were displayed in Ref.~\cite{romao}
as linear combinations of the dispersive parts
of various Passarino--Veltman functions (PVF)~\cite{pv}.
The PVF may be computed,
for instance,
by using the software {\tt LoopTools}~\cite{LT1,LT2}.
However,
it may be more convenient to present formulas
for the functions that do not involve the PVF
and that may be more immediately written in a code.
That's what we do here.
The functions are:
\bs
\label{cjofpd}
\ba
k \left( Q, I, J \right) &=&
\frac{1}{3} - \frac{I + J}{4 Q} - \frac{\left( I - J \right)^2}{2 Q^2}
+ \frac{1}{4 Q} \left[ \frac{\left( I - J \right)^3}{Q^2}
  - \frac{I^2 + J^2}{I - J} \right] \ln{\frac{I}{J}}
\no & &
+ \left[ - I - J + \frac{\left( I + J \right)^2}{Q}
  + \frac{\left( I - J \right)^2 \left( I + J \right)}{Q^2} 
  - \frac{\left( I - J \right)^4}{Q^3} \right]
\frac{f \left( Q, I, J \right)}{4},
\label{6a}
\\
j \left( Q, I, J \right) &=&
- 2 + \left[ \frac{I - J}{Q} - \frac{I + J}{2 \left( I - J \right)} \right]
\ln{\frac{I}{J}}
\no & &
+ \left[ - \frac{Q}{2} + \frac{3 \left( I + J \right)}{2}
  - \frac{\left( I - J \right)^2}{Q} \right] f \left( Q, I, J \right),
\label{6b}
\\
g \left( Q, I, J \right) &=&
- \frac{\mathrm{div}}{3} + \frac{1}{6} \left( \ln{\frac{I}{\mu^2}}
+ \ln{\frac{J}{\mu^2}} \right)
- \frac{5}{9} + \frac{I + J}{12 Q} + \frac{\left( I - J \right)^2}{6 Q^2}
\no & &
+ \frac{1}{4 Q} \left[ \frac{I^2 + J^2}{I - J}
  - \frac{\left( I - J \right)^3}{3 Q^2} \right] \ln{\frac{I}{J}}
\no & &
+ \left[ - 2 Q + 5 \left( I + J \right)
  - \frac{3 I^2 + 3 J^2 + 2 I J}{Q}
  \right. \no & & \left.
  - \frac{\left( I + J \right) \left( I - J \right)^2}{Q^2}
  + \frac{\left( I - J \right)^4}{Q^3} \right]
\frac{f \left( Q, I, J \right)}{12},
\label{6c} \\
\hat g \left( Q, I, J \right) &=&
1 + \frac{1}{2} \left( \frac{I + J}{I - J}
  - \frac{I - J}{Q} \right) \ln{\frac{I}{J}}
\no & &
+ \left[ \frac{Q}{2} - I - J + \frac{\left( I - J \right)^2}{2 Q} \right]
f \left( Q, I, J \right),
\label{6d} \\
l \left( Q, I \right) &=& - \frac{5}{9} - \frac{4 I}{3 Q}
+ \left( - \frac{Q}{6} + \frac{I}{3} + \frac{4 I^2}{3 Q} \right)
f \left( Q, I, I \right),
\label{6e} \\
h \left( I \right) &=& \frac{\mathrm{div}}{3} - \frac{1}{3}\,
\ln{\frac{I}{\mu^2}},
\label{6f} \\
t \left( I, J \right) &=& \frac{I + J}{4} \left[ \mathrm{div}
  - \frac{1}{2} \left( \ln{\frac{I}{\mu^2}} + \ln{\frac{J}{\mu^2}} \right)
  \right]
+ \frac{I + J}{8}
- \frac{I^2 + J^2}{8 \left( I - J \right)}\, \ln{\frac{I}{J}},
\label{6g} \\
\hat t \left( I, J \right) &=& \mathrm{div}
- \frac{1}{2} \left( \ln{\frac{I}{\mu^2}} + \ln{\frac{J}{\mu^2}} \right)
+ 1 - \frac{I + J}{2 \left( I - J \right)}\, \ln{\frac{I}{J}}.
\label{6h}
\ea
\es
In Eqs.~\eqref{6a}--\eqref{6e},
\be
f \left( Q, I, J \right) = \left\{ \begin{array}{l}
  \displaystyle{\frac{1}{\sqrt{\Delta}}\,
    \ln{\frac{I + J - Q + \sqrt{\Delta}}{I + J - Q - \sqrt{\Delta}}}\
    \Leftarrow\ \Delta > 0}, \\*[3mm]
  \displaystyle{\frac{2}{\sqrt{- \Delta}} \left(
    \arctan{\frac{Q + I - J}{\sqrt{- \Delta}}}
    + \arctan{\frac{Q + J - I}{\sqrt{- \Delta}}} \right)\
    \Leftarrow\ \Delta < 0}, \\*[3mm]
  \displaystyle{\frac{1}{\sqrt{I J}}\ \Leftarrow \sqrt{Q} = \left|
    \sqrt{I} - \sqrt{J} \right|}, \\*[3mm]
  \displaystyle{\frac{- 1}{\sqrt{I J}}\ \Leftarrow \sqrt{Q} =
    \sqrt{I} + \sqrt{J}},
\end{array} \right.
\ee
where
\be
\Delta = Q^2 - 2 Q \left( I + J \right) + \left( I - J \right)^2.
\ee
Equations~\eqref{6a}--\eqref{6d},
\eqref{6g},
and~\eqref{6h} have been written assuming $I \neq J$.
It is easy to find the fitting expressions for $I = J$:
\bs
\label{ogfpfd}
\ba
k \left( Q, I, I \right) &=& \frac{1}{3} - \frac{I}{Q}
+ I \left( \frac{I}{Q} - \frac{1}{2} \right) f \left( Q, I, I \right),
\label{8a} \\
j \left( Q, I, I \right) &=& - 3
+ \left( 3 I - \frac{Q}{2} \right) f \left( Q, I, I \right),
\label{8b} \\
g \left( Q, I, I \right) &=& - \frac{\mathrm{div}}{3}
+ \frac{1}{3}\, \ln{\frac{I}{\mu^2}} - \frac{5}{9} + \frac{2 I}{3 Q}
- \left( Q - 5 I + \frac{4 I^2}{Q} \right)
\frac{f \left( Q, I, I \right)}{6},
\label{8c} \\
\hat g \left( Q, I, I \right) &=& 2 + \left( Q - 4 I \right)
\frac{f \left( Q, I, I \right)}{2},
\label{8d} \\
t \left( I, I \right) &=& \frac{I}{2} \left( \mathrm{div}
- \ln{\frac{I}{\mu^2}} \right),
\label{8e} \\
\hat t \left( I, I \right) &=& \mathrm{div} - \ln{\frac{I}{\mu^2}}.
\label{8f}
\ea
\es

Some Eqs.~\eqref{cjofpd} and~\eqref{ogfpfd}
depend on a dimensionless divergent quantity `div'
and on an arbitrary mass parameter $\mu$;
those two quantities are supposed to disappear from the formulas
for any physical quantity like the OPs.
We will soon see the way that happens in practice.

When the masses of the New Physics particles
are much larger than the Fermi scale one may use,
instead of expressions~\eqref{6a}--\eqref{6e},
their approximations for $Q \ll I, J$.
Thus,
\bs
\ba
k \left( Q, I, J \right) &=&
\left\{ \begin{array}{lcl}
\displaystyle{Q \left[ \frac{\left( I + J \right)
    \left( 8 I J - I^2 - J^2 \right)}{8 \left( I - J \right)^4}
  - \frac{3 I^2 J^2}{2 \left( I - J \right)^5}\,
  \ln{\frac{I}{J}} \right]
  + \mathrm{O} \left( Q^2 \right)} &\Leftarrow& I \neq J,
\\*[3mm]
\displaystyle{- \frac{Q}{20 I} + \mathrm{O} \left( Q^2 \right)}
&\Leftarrow& I = J,
\end{array} \right.
\\*[5mm]
\frac{j \left( Q, I, J \right)}{Q} &=& \left\{ \begin{array}{lcl}
  \displaystyle{Q \left[
  \frac{10 I J + I^2 + J^2}{6 \left( I - J \right)^4}
  - \frac{I J \left( I + J \right)}{\left( I - J \right)^5}\,
  \ln{\frac{I}{J}} \right] + \mathrm{O} \left( Q^2 \right)}
  &\Leftarrow& I \neq J,
  \\*[3mm]
  \displaystyle{\frac{Q}{60 I^2} + \mathrm{O} \left( Q^2 \right)}
  &\Leftarrow& I = J,
  \end{array} \right.
\\*[5mm]
g \left( Q, I, J \right) &=&
\overline{g} \left( I, J \right) + k \left( Q, I, J \right),
%%%%% I ADDED A LABEL:
\label{pipo1}
\\
\frac{\hat g \left( Q, I, J \right)}{Q} &=&
\overline{\hat g} \left( I, J \right) + \frac{j \left( Q, I, J \right)}{Q},
%%%%% I ADDED A LABEL:
\label{pipo2}
\\
l \left( Q, I \right) &=&
- \frac{Q}{15\, I} + \mathrm{O} \left( Q^2 \right),
%%%%% I ADDED A LABEL:
\label{mflspdtk}
\ea
\es
where
\be
\overline{g} \left( I, J \right) =
\left\{ \begin{array}{lcl}
\displaystyle{- \frac{\mathrm{div}}{3}
+ \frac{1}{6} \left( \ln{\frac{I}{\mu^2}} + \ln{\frac{J}{\mu^2}} \right)
+ \frac{8 I J - I^2 - J^2}{9 \left( I - J \right)^2}} & &
\\*[5mm]
\displaystyle{\hspace*{25mm}
  + \frac{I^3 + J^3 - 3 I^2 J - 3 I J^2}{6 \left( I - J \right)^3}\,
\ln{\frac{I}{J}}} &\Leftarrow& I \neq J,
\\*[5mm]
\displaystyle{- \frac{\mathrm{div}}{3} + \frac{1}{3}\, \ln{\frac{I}{\mu^2}}
  + \frac{1}{6}}
&\Leftarrow& I = J,
\end{array} \right.
\label{g000}
\ee
and
\be
\overline{\hat g} \left( I, J \right) =
\left\{ \begin{array}{lcl}
\displaystyle{\frac{I + J}{2 \left( I - J \right)^2}
  - \frac{I J}{\left( I - J \right)^3}\, \ln{\frac{I}{J}}}
&\Leftarrow& I \neq J,
\\*[5mm]
\displaystyle{\frac{1}{6 I}} &\Leftarrow& I = J. \end{array} \right.
\label{hatg000}
\ee

\section{Mixing matrices}
\label{sec:mixing}

We put together in a set all the fermions
that have the same chirality $E$
($E$ may be either $L$---left---or $R$---right) and the same colour.
If there are in the NPM any other non-$SU(2) \times U(1)$
conserved quantum numbers,
then all the fermions in each set should have the same values
of those quantum numbers too.
Moreover,
all the fermions in each set
must have electric charges that differ among themselves
by \emph{integer} numbers;
this means that,
if any two fermions have electric charges that differ between themselves
through a non-integer,
then those two fermions must be placed in \emph{different} sets.
We emphasize that different sets must be treated separately,
because they give separate contributions to each OP,
just as new scalars in an NPM give separate contributions to the OPs
from new fermions in the NPM.

%It is possible that the NPM under consideration
%includes some fermions that mix with the SM fermions.
%This happens, in particular,
%when the NPM includes either colour triplets
%with electric charge either $2/3$ or $-1/3$,
%or colour singlets with electric charge either $0$ or $\pm 1$.
%If this happens,
%then one must follow the recipe below
%to compute firstly the OPs in the NPM
%and secondly the OPs in the SM,
%and then subtract the second OPs
%from the first corresponding OPs,
%\ie\ one must compute
%$O = O^\mathrm{NPM} - O^\mathrm{SM}$
%for each oblique parameter $O$.

We consider in turn each set of fermions with chirality $E$.
In the set,
the raising operator of weak isospin,
\viz\ $T_+$,
is represented by a matrix that we name
$M_E \left/ \sqrt{2} \right.$.\footnote{The denominator $\sqrt{2}$
is purely conventional.}
The lowering operator of weak isospin,
\ie\ $T_-$,
is the Hermitian conjugate of $T_+$;
therefore,
it is represented by the matrix $M_E^\dagger \left/ \sqrt{2} \right.$.
Finally,
the third component of weak isospin is
\be
T_3 = \left[ T_+,\, T_- \right]
\ee
and is represented by the matrix $\left. H_E \right/ \! 2$,\footnote{The
denominator $2$ is just a convention.}
where
\be
\label{he}
H_E = \left[ M_E,\, M_E^\dagger \right].
\ee

We must take into account the weak-isospin commutation relation
\be
\label{uvifo}
\left[ T_3,\, T_+ \right] = T_+.
\ee
Since,
as written in the previous paragraph,
$T_+ \mapsto M_E \left/ \sqrt{2} \right.$
and $T_3 \mapsto \left. \left[ M_E,\, M_E^\dagger \right] \right/ 2$,
Eq.~\eqref{uvifo} implies
\be
\label{jvifo}
M_E = M_E M_E^\dagger M_E - \frac{1}{2} \left(
M_E^2 M_E^\dagger + M_E^\dagger M_E^2 \right).
\ee
Equation~\eqref{jvifo} implies
\bs
\label{s}
\ba
\mathrm{tr} \left( M_E M_E^\dagger \right)
&=& \mathrm{tr} \left( M_E M_E^\dagger M_E M_E^\dagger \right)
- \mathrm{tr} \left( M_E^2 {M_E^\dagger}^2 \right)
\\ &=& \frac{\mathrm{tr} \left( H_E^2 \right)}{2}.
\ea
\es
Equation~\eqref{s} is separately valid for each set of fermions;
in particular,
it is valid for both $E = L$ and $E = R$.

We place the fermions of each set in a column vector,
ordering them by decreasing electric charges.
This means that the electric-charge operator
is represented by the square matrix\footnote{We implicitly assume
that the electric charges of the left-handed fermions
are the same as those of the right-handed fermions,
so that all the fermions may acquire a Dirac mass.}
\be
\label{qqq1}
Q = \left( \begin{array}{cccc}
  Q_1 \times \mathbf{1}_{q_1} &
  \mathbf{0}_{q_1 \times q_2} & \mathbf{0}_{q_1 \times q_3} & \cdots \\
  \mathbf{0}_{q_2 \times q_1} & Q_2 \times \mathbf{1}_{q_2} &
  \mathbf{0}_{q_2 \times q_3} & \cdots \\
  \mathbf{0}_{q_3 \times q_1} & \mathbf{0}_{q_3 \times q_2} &
  Q_3 \times \mathbf{1}_{q_3} & \cdots \\
  \vdots & \vdots & \vdots & \ddots
\end{array} \right),
\ee
where $\mathbf{0}_{m \times n}$ denotes the $m \times n$ null matrix,
$\mathbf{1}_n$ denotes the $n \times n$ unit matrix,
$q_n$ is the number of fermions in the set that have electric charge $Q_n$,
and
\be
\label{jgioc}
Q_1 - Q_2 = Q_2 - Q_3 = \cdots = 1.
\ee
When one adopts this ordering of the fermions in a set we see that,
since $T_+$ connects the fermions of a given electric charge
to the fermions with one unit less of electric charge,
one must have
\bs
\label{jvifod}
\ba
M_E &=& \left( \begin{array}{ccccc}
  \mathbf{0}_{q_1 \times q_1} &
  M_{E 1} & \mathbf{0}_{q_1 \times q_3} &
  \mathbf{0}_{q_1 \times q_4} & \cdots \\
  \mathbf{0}_{q_2 \times q_1} & \mathbf{0}_{q_2 \times q_2} &
  M_{E 2} & \mathbf{0}_{q_2 \times q_4} & \cdots \\
  \mathbf{0}_{q_3 \times q_1} & \mathbf{0}_{q_3 \times q_2} &
  \mathbf{0}_{q_3 \times q_3} & M_{E 3} & \cdots \\
  \vdots & \vdots & \vdots & \vdots & \ddots
\end{array} \right),
\\
M_E^\dagger &=& \left( \begin{array}{ccccc}
  \mathbf{0}_{q_1 \times q_1} & \mathbf{0}_{q_1 \times q_2} &
  \mathbf{0}_{q_1 \times q_3} & \mathbf{0}_{q_1 \times q_4} &
  \cdots \\
  M_{E 1}^\dagger & \mathbf{0}_{q_2 \times q_2} &
  \mathbf{0}_{q_2 \times q_3} & \mathbf{0}_{q_2 \times q_4} &
  \cdots \\
  \mathbf{0}_{q_3 \times q_1} & M_{E 2}^\dagger &
  \mathbf{0}_{q_3 \times q_3} & \mathbf{0}_{q_3 \times q_4} &
  \cdots \\
  \vdots & \vdots & \vdots & \vdots & \ddots
\end{array} \right),
\ea
\es
where $M_{E n}$ is a $q_n \times q_{n+1}$ matrix.\footnote{For instance,
it is well known that for a doublet of $SU(2)$
\be
M_E = \left( \begin{array}{cc} 0 & 1 \\ 0 & 0 \end{array} \right),
\ee
while for a triplet of $SU(2)$
\be
M_E = \left( \begin{array}{ccc} 0 & \sqrt{2} & 0 \\ 0 & 0 & \sqrt{2} \\
0 & 0 & 0 \end{array} \right).
\ee
}
Then,
\be
\label{cjifof}
H_E = \left( \begin{array}{cccc}
  M_{E 1} M_{E 1}^\dagger &
  \mathbf{0}_{q_1 \times q_2} & \mathbf{0}_{q_1 \times q_3} & \cdots \\
  \mathbf{0}_{q_2 \times q_1} &
  M_{E 2} M_{E 2}^\dagger - M_{E 1}^\dagger M_{E 1} &
  \mathbf{0}_{q_2 \times q_3} & \cdots \\
  \mathbf{0}_{q_3 \times q_1} & \mathbf{0}_{q_3 \times q_2} &
  M_{E 3} M_{E 3}^\dagger - M_{E 2}^\dagger M_{E 2} & \cdots \\
  \vdots & \vdots & \vdots & \ddots
\end{array} \right).
\ee
The $Z^0$ boson couples to $T_3 - Q s_W^2$.
Since $T_3 \mapsto \left. H_E \right/ 2$,
it is convenient to define the matrix $F_E$ through
\be
\label{F}
F_E = H_E - 2 s_W^2 Q,
\ee
where $Q$ is the diagonal,
real matrix in Eq.~\eqref{qqq1}.
The matrices $F_E$ are Hermitian just as the matrices $H_E$.

Using Eqs.~\eqref{jvifod} we see that
\be
\mathrm{tr} \left( M_E M_E^\dagger \right) =
\mathrm{tr} \left( M_{E 1} M_{E 1}^\dagger \right)
+ \mathrm{tr} \left( M_{E 2} M_{E 2}^\dagger \right)
+ \cdots.
\ee
Also,
using Eqs.~\eqref{qqq1} and~\eqref{cjifof},
\be
\mathrm{tr} \left( Q H_E \right) =
\left( Q_1 - Q_2 \right)
\mathrm{tr} \left( M_{E 1} M_{E 1}^\dagger \right)
+ \left( Q_2 - Q_3 \right)
\mathrm{tr} \left( M_{E 2} M_{E 2}^\dagger \right)
+ \cdots.
\ee
Utilizing Eq.~\eqref{jgioc} we then conclude that
\be
\label{u}
\mathrm{tr} \left( Q H_E \right)
  = \mathrm{tr} \left( M_E M_E^\dagger \right).
\ee
Equations~\eqref{s} and~\eqref{u}
are crucial to demonstrate the finiteness of the oblique parameters $S$ and $U$.
Notice that those two equations depend
neither on the masses of the fermions
nor on the way that those masses are generated.

Notice that in this formalism we do not mention the hypercharge $Y$ at all.
In a weak basis each fermion has a well-defined $T_3$ and a well-defined $Y$.
In the physical basis that we utilize this is not so:
each physical fermion may be the superposition of various components
with different $T_3$ and different $Y$.
On the other hand,
$Q = T_3 + Y$ has a well-defined value $Q_f$ for each physical fermion $f$.

Using the covariant derivative~\cite{book}
\be
D_\mu = \partial_\mu
+ i e A_\mu Q
- i\, \frac{e}{s_W} \left( W^+_\mu T_+ + W^-_\mu T_- \right)
- i\, \frac{e}{s_W c_W}\, Z_\mu \left( T_3 - Q s_W^2 \right),
\ee
where $e = \sqrt{4 \pi \alpha}$ is the electric-charge unit,
we may now write the gauge-kinetic Lagrangian for the fermions $f_E$
in a set:
\ba
\mathcal{L}_\mathrm{gk} &=& \frac{i}{2}\, \sum_f \left[
  \bar f_E\, \gamma^\mu \left( \partial_\mu f_E \right)
  - \left( \partial_\mu \bar f_E \right) \gamma^\mu\, f_E \right]
\no & &
- e A_\mu\, \sum_f Q_f\, \bar f_E \gamma^\mu f_E
+ \frac{e}{2 s_W c_W}\ Z_\mu\, \sum_{f, f^\prime}
\left( F_E \right)_{f f^\prime} \bar f_E \gamma^\mu f^\prime_E
\no & &
+ \frac{e}{\sqrt{2} s_W}\, \sum_{f, f^\prime} \left[
  W_\mu^+ \left( M_E \right)_{f f^\prime}
  + W_\mu^- \left( M_E^\dagger \right)_{f f^\prime}
  \right] \bar f_E \gamma^\mu f^\prime_E.
\ea

\section{Formulas for the OPs}
\label{sec:ops}

Using the computations in Ref.~\cite{romao},
we are now in a position to write the formulas for the various OPs.

\paragraph{The parameters $V$ and $W$:} One has
\bs
\label{vw}
\ba
V &=& \frac{1}{8 \pi s_W^2 c_W^2}\, \sum_{f, f^\prime}\,
\mathcal{F} \left[ \left( F_L \right)_{f f^\prime}, \left( F_R \right)_{f f^\prime},
  m_Z^2, m_f^2, m_{f^\prime}^2 \right],
\label{V} \\
W &=& \frac{1}{4 \pi s_W^2}\, \sum_{f, f^\prime}\,
\mathcal{F} \left[ \left( M_L \right)_{f f^\prime}, \left( M_R \right)_{f f^\prime},
  m_W^2, m_f^2, m_{f^\prime}^2 \right],
  \label{W}
\ea
\es
where the sum runs over all the fermions $f$ and $f^\prime$ in a set,
$m_f$ and $m_{f^\prime}$ are the masses of $f$ and $f^\prime$,
respectively,
and
\be
\label{mathcalf}
\mathcal{F} \left( x, y, Q, I, J \right) = \left( \left| x \right|^2
+ \left| y \right|^2 \right) k \left( Q, I, J \right)
- 2\, \mathrm{Re} \left( x y^\ast \right)\,
\frac{j \left( Q, I, J \right)}{Q}\ \sqrt{I J}.
\ee
It is worth pointing out that in Eq.~\eqref{V},
whenever $f \neq f^\prime$,
there are \emph{two equal terms} in the sum,
because the matrices $F_L$ and $F_R$ are Hermitian and
\be
\mathcal{F} \left( x, y, Q, I, J \right)
= \mathcal{F} \left( x^\ast, y^\ast, Q, J, I \right).
\ee

\paragraph{The parameter $X$:} One has
\be
\label{x}
X = \frac{1}{4 \pi}\, \sum_f\, Q_f \left( F_L + F_R \right)_{ff}\,
l \left( m_Z^2, m_f^2 \right).
\ee

\paragraph{The parameters $S$ and $U$:} One has
\bs
\label{su}
\ba
S &=& \frac{1}{2 \pi}\, \sum_{f, f^\prime}\,
\mathcal{G} \left[ \left( H_L \right)_{f f^\prime}, \left( H_R \right)_{f f^\prime},
  m_Z^2, m_f^2, m_{f^\prime}^2 \right]
\no & &
+ \frac{1}{\pi}\, \sum_f\, Q_f \left( H_L + H_R \right)_{ff}\,
h \left( m_f^2 \right)
\no & &
+ \frac{2 s_W^2}{\pi}\, \sum_f\, Q_f \left[2 Q_f s_W^2
  - \left( H_L + H_R \right)_{ff} \right] l \left( m_Z^2, m_f^2 \right),
\label{sss}
\\
U &=& - S
+ \frac{1}{\pi}\, \sum_{f, f^\prime}\,
\mathcal{G} \left[ \left( M_L \right)_{f f^\prime}, \left( M_R \right)_{f f^\prime},
  m_W^2, m_f^2, m_{f^\prime}^2 \right]
\no & &
+ \frac{1}{\pi}\, \sum_f\, Q_f \left( H_L + H_R \right)_{ff}\,
h \left( m_f^2 \right),
\label{uuu}
\ea
\es
where
\be
\mathcal{G} \left( x, y, Q, I, J \right) = \left( \left| x \right|^2
+ \left| y \right|^2 \right) g \left( Q, I, J \right)
- 2\, \mathrm{Re} \left( x y^\ast \right)\,
\frac{\hat g \left( Q, I, J \right)}{Q}\ \sqrt{I J}.
\label{ggg}
\ee
Note that in Ref.~\cite{romao} the function $\mathcal{G}$ was defined
with the opposite sign.

\paragraph{Cancellation of the divergence in $S$:} We remind that,
according to Eqs.~\eqref{cjofpd},
\bs
\label{divs}
\ba
g \left( Q, I, J \right) &=& - \frac{\widetilde{\mathrm{div}}}{3}
+ \mathrm{finite,}\ \mu\mbox{-independent\ terms},
\\
h \left( I \right) &=& \frac{\widetilde{\mathrm{div}}}{3}
+ \mathrm{finite,}\ \mu\mbox{-independent\ terms},
\ea
\es
and the functions $\hat g \left( Q, I, J \right)$ and $l \left( Q, I \right)$
do not contain $\widetilde{\mathrm{div}}$,
where $\widetilde{\mathrm{div}} \equiv \mathrm{div} + \ln{\mu^2}$
includes both the divergent quantity `$\mathrm{div}$'
and the arbitrary mass $\mu$.
From Eqs.~\eqref{sss},
\eqref{ggg},
and~\eqref{divs} one sees that
\ba
S &=& - \frac{\widetilde{\mathrm{div}}}{6 \pi}\, \sum_{f, f^\prime}\,
\left[ \left| \left( H_L \right)_{f f^\prime} \right|^2
+ \left| \left( H_R \right)_{f f^\prime} \right|^2 \right]
+ \frac{\widetilde{\mathrm{div}}}{3 \pi}\,
\sum_f\, Q_f \left( H_L + H_R \right)_{ff}
\no & &
+ \mathrm{finite,}\ \mu\mbox{-independent\ terms}.
\ea
But $H_L$ and $H_R$ are Hermitian matrices,
therefore
\ba
S &=& \frac{\widetilde{\mathrm{div}}}{6 \pi} \left\{ - \mathrm{tr}
\left[ \left( H_L \right)^2 + \left( H_R \right)^2 \right]
+ 2\, \mathrm{tr} \left[ Q \left( H_L + H_R \right) \right] \right\}
\no & &
+ \mathrm{finite,}\ \mu\mbox{-independent\ terms}.
\label{jvifoo}
\ea
The terms in Eq.~\eqref{jvifoo} proportional to $\widetilde{\mathrm{div}}$
vanish because of Eqs.~\eqref{s} and~\eqref{u}
(actually,
they vanish separately for $E=L$ and $E=R$).
Thus,
$S$ is both finite and $\mu$-independent.

\paragraph{Cancellation of the divergence in $U$:} Since $S$ is
$\widetilde{\mathrm{div}}$-independent,
Eqs.~\eqref{uuu},
\eqref{ggg},
and~\eqref{divs} produce
\ba
U &=& \frac{\widetilde{\mathrm{div}}}{3 \pi} \left\{
- \sum_{f, f^\prime} \left[ \left| \left( M_L \right)_{f f^\prime} \right|^2
  + \left| \left( M_R \right)_{f f^\prime} \right|^2 \right]
+ \sum_f Q_f \left( H_L + H_R \right)_{ff} \right\}
\no & &
\no & &
+ \mathrm{finite,}\ \mu\mbox{-independent\ terms}.
\ea
Therefore,
\ba
U &=& \frac{\widetilde{\mathrm{div}}}{3 \pi} \left\{ - \mathrm{tr}
\left( M_L M_L^\dagger + M_R M_R^\dagger \right)
+ \mathrm{tr} \left[ Q \left( H_L + H_R \right) \right] \right\}
\no & &
+ \mathrm{finite,}\ \mu\mbox{-independent\ terms}.
\label{78}
\ea
The $\widetilde{\mathrm{div}}$-dependent terms
in Eq.~\eqref{78} vanish because of Eq.~\eqref{u}.
Thus,
$U$ is both finite and $\mu$-independent.

\paragraph{The parameter $T$:} One has
\ba
T &=& \frac{1}{4 \pi s_W^2 m_W^2}\, \sum_{f, f^\prime}\, \left\{
2\, \mathcal{H} \left[ \left( M_L \right)_{f f^\prime},
  \left( M_R \right)_{f f^\prime}, m_f^2, m_{f^\prime}^2 \right]
\right. \no & & \left.
- \mathcal{H} \left[ \left( H_L \right)_{f f^\prime},
  \left( H_R \right)_{f f^\prime}, m_f^2, m_{f^\prime}^2 \right] \right\},
\label{79}
\ea
where
\be
\label{uh}
\mathcal{H} \left( x, y, I, J \right)
= \left( \left| x \right|^2 + \left| y \right|^2 \right)
t \left( I, J \right)
- \mathrm{Re} \left( x y^\ast \right) \sqrt{I J}\ \, \hat t \left( I, J \right).
\ee
We note that,
because of Eqs.~\eqref{8e} and~\eqref{8f},
\be
\label{mfjdoo}
\mathcal{H} \left( x, y, I, I \right)
= \frac{I}{2}\ \hat t \left( I, I \right)\, \left| x - y \right|^2.
\ee
In the second line of Eq.~\eqref{79}
there are two equal terms in the sum whenever $f \neq f^\prime$,
because
\be
\mathcal{H} \left( x, y, I, J \right)
= \mathcal{H} \left( x^\ast, y^\ast, J, I \right).
\ee

Note that
\be
2\, \mathcal{H} \left( x, y, I, J \right)
- \mathcal{H} \left( x, y, I, I \right)
- \mathcal{H} \left( x, y, J, J \right)
= \left( \left| x \right|^2 + \left| y \right|^2 \right)
\frac{\theta_+ \left( I, J \right)}{4}
+ \mathrm{Re} \left( x y^\ast \right)
\frac{\theta_- \left( I, J \right)}{2},
\ee
where $\theta_+ \left( I, J \right)$ and $\theta_- \left( I, J \right)$
are the functions that were
defined in Equations (12) and (13) of Ref.~\cite{silva}.

\paragraph{Cancellation of the divergence in $T$:} Because of
Eqs.~\eqref{6g} and~\eqref{6h},
\bs
\ba
t \left( I, J \right) &=& \frac{I + J}{4}\ \widetilde{\mathrm{div}}
+ \mathrm{finite,}\ \mu\mbox{-independent\ terms},
\\
\hat t \left( I, J \right) &=& \widetilde{\mathrm{div}}
+ \mathrm{finite,}\ \mu\mbox{-independent\ terms}.
\ea
\es
Therefore,
\bs
\ba
T &=& \frac{\widetilde{\mathrm{div}}}{16 \pi s_W^2 m_W^2}\,
\sum_{f, f^\prime}\, \left\{
2 \left[ \left| \left( M_L \right)_{f f^\prime} \right|^2
  + \left| \left( M_R \right)_{f f^\prime} \right|^2 \right]
\left( m_f^2 + m_{f^\prime}^2 \right)
\right. \no & &
- 8\, \mathrm{Re} \left[
  \left( M_L \right)_{f f^\prime} \left( M_R^\ast \right)_{f f^\prime} \right]
m_f m_{f^\prime}
- \left[ \left| \left( H_L \right)_{f f^\prime} \right|^2
  + \left| \left( H_R \right)_{f f^\prime} \right|^2 \right]
\left( m_f^2 + m_{f^\prime}^2 \right)
\no & & \left.
+ 4\, \mathrm{Re} \left[ \left( H_L \right)_{f f^\prime}
  \left( H_R^\ast \right)_{f f^\prime} \right]
m_f m_{f^\prime}
\right\}
+ \mathrm{finite,}\ \mu\mbox{-independent\ terms}
\\ &=& \frac{\widetilde{\mathrm{div}}}{8 \pi s_W^2 m_W^2} \left\{
\mathrm{tr} \left[ \left( M_L M_L^\dagger + M_L^\dagger M_L
  + M_R M_R^\dagger + M_R^\dagger M_R \right) M^2 \right]
\right. \no & &
- 2\, \mathrm{tr} \left( M_L M M_R^\dagger M + M_R M M_L^\dagger M \right)
- \mathrm{tr} \left[ \left( H_L^2 + H_R^2 \right) M^2 \right]
\no & & \left.
+ 2\, \mathrm{tr} \left( H_L M H_R M \right)
\vphantom{\mathrm{tr} \left[ \left( M_L M_L^\dagger \right) \right]}
\right\}
+ \mathrm{finite,}\ \mu\mbox{-independent\ terms},
\ea
\es
where $M$ is the mass matrix of the fermions.
Thus,
$T$ is finite and $\mu$-independent if
\bs
\label{fjvfio}
\ba
\mathrm{tr}
\left[ \left( M_L M_L^\dagger + M_L^\dagger M_L - H_L^2 \right) M^2 \right]
& & \label{tra} \\
+ \mathrm{tr}
\left[ \left( M_R M_R^\dagger + M_R^\dagger M_R - H_R^2 \right) M^2 \right]
& & \label{trb} \\
+ 2\ \mathrm{tr}
\left( H_L M H_R M - M_L M M_R^\dagger M - M_R M M_L^\dagger M \right)
&=& 0.
\label{jvifo0}
\ea
\es
The oblique parameter $T$ is \emph{not} automatically finite,
contrary to what happens with $S$ and $U$.
This should not surprise us.
It is well known that $T$ is divergent
when the NPM does not obey Eq.~\eqref{ibvugpgf} at the tree level.
In our case,
the fermions may get masses either through bare mass terms,
if they are in vector-like representations of $SU(2) \times U(1)$,
or through their Yukawa couplings to neutral-scalar fields
and the VEVs of those fields.
Now,
the VEVs may cause a violation
of Eq.~\eqref{ibvugpgf} if the neutral-scalar fields
do not feature $J \left( J + 1 \right) = 3 Y^2$.
If the fermion mass matrix $M$ implicitly requires some scalar fields
to have disallowed VEVs,
then Eq.~\eqref{fjvfio} does not hold
and $T$ is divergent.\footnote{The fact
that $T$ may turn out divergent when one adds fermions to the SM
and one gives arbitrary masses to those fermions
had already been pointed out in Ref.~\cite{erro}.}

\section{One vector-like multiplet}
\label{sec:one}

We consider in this section the simple case
of one vector-like multiplet of fermions with isospin $J$ and hypercharge $Y$.
All the $n = 2 J + 1$
components of the multiplet have the same (bare) mass $m$,
because there are,
in general,
no Yukawa couplings that can generate different masses
for the different components of the multiplet.
So,
the only variables in this model are $m$ and $Y$,
which are continuous,
and $n$,
which is an integer.

The $n \times n$ matrices $M_L$ and $M_R$ are equal and they are given by
\be
\left( M_L \right)_{rc}
= \left( M_R \right)_{rc}
= \delta_{c,r+1} \sqrt{r \left( n - r \right)},
\label{j1}
\ee
where the sub-index $r$ stands for ``row''
and the sub-index $c$ stands for ``column'' of a matrix.
The $n \times n$ matrices $H_L$ and $H_R$ are equal and they are given by
\be
\left( H_L \right)_{rc}
= \left( H_R \right)_{rc}
= \delta_{c,r} \left( n + 1 - 2 r \right).
\label{j2}
\ee
The electric-charge matrix is given by
\be
Q_{rc} = \delta_{c,r}\, \frac{n + 1 - 2 r + 2 Y}{2}.
\label{j3}
\ee
The $n \times n$ matrices $F_L$ and $F_R$ are equal and they are given by
\be
\left( F_L \right)_{rc}
= \left( F_R \right)_{rc}
= \delta_{c,r} \left[ \left( n + 1 - 2 r \right) c_W^2 - 2 Y s_W^2 \right].
\ee

\begin{figure}[ht]
\begin{center}
\includegraphics[width=1.0\textwidth]{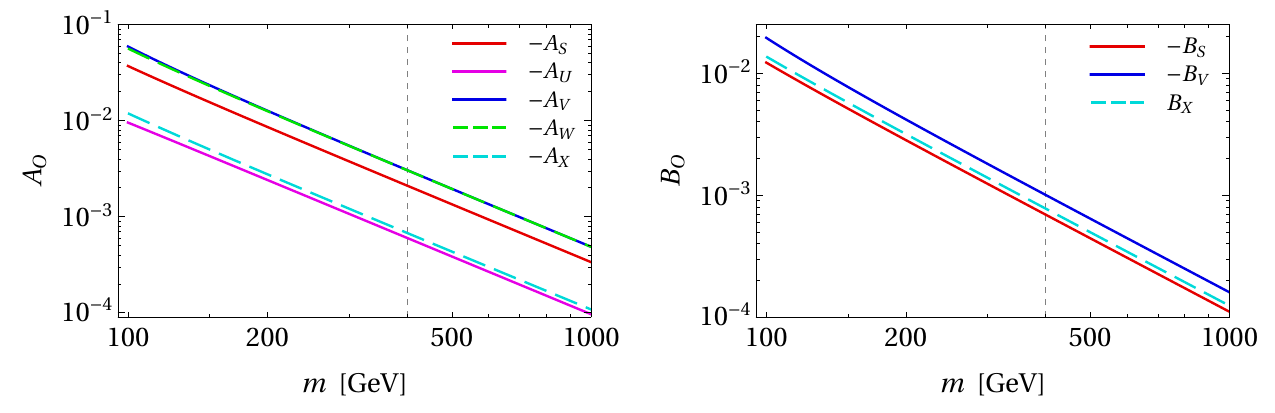}
\end{center}
%\vspace{-6mm}
\caption{The coefficients $A_O$ and $B_O$ as functions of $m$ according to 
Eqs.~\eqref{eq_AO} and~\eqref{eq_BO}. 
The dashed vertical line indicates the benchmark point value $m=400$~GeV.
For large $m$, all the coefficients $A_O$ and $B_O$ vary as $m^{-p}$
with $p$ very close to 2.
%The coefficients obey following approximations: $A_S \approx -365\, m^{-2.01}$, $A_U \approx -96\, m^{-2.0}$, $A_V \approx A_W \approx -570\, m^{-2.02}$, $A_X \approx -118\, m^{-2.01}$, $B_S \approx -123\, m^{-2.01}$, $B_V \approx -188\, m^{-2.02}$ and $B_X \approx 140\, m^{-2.015}$.
}
\label{1multp_ABo_vs_m}
\end{figure}

Because of Eq.~\eqref{mfjdoo}
and of the equalities between the matrices $M_L$ and $M_R$
and between the matrices $F_L$ and $F_R$,
the oblique parameter $T$ vanishes.
For the remaining OPs $O = S, U, V, W, X$
we obtain the general expression
\be
O = \frac{n}{\pi} \left( A_O\, \frac{n^2 - 1}{3} + B_O Y^2 \right),
\ee
where the coefficients $A_O$ and $B_O$ depend neither on $n$ nor on $Y$;
they only depend on $m$:
\bs
\label{eq_AO}
\ba
A_S &=& g \left( m_Z^2, m^2, m^2 \right)
- \frac{m^2}{m_Z^2}\ \hat g \left( m_Z^2, m^2, m^2 \right)
\no & &
+ h \left( m^2 \right)
- s_W^2 \left( 2 - s_W^2 \right)\, l \left( m_Z^2, m^2 \right)
\\
A_U &=& g \left( m_W^2, m^2, m^2 \right) - g \left( m_Z^2, m^2, m^2 \right)
\no & &
+ m^2 \left[ \frac{\hat g \left( m_Z^2, m^2, m^2 \right)}{m_Z^2}
  - \frac{\hat g \left( m_W^2, m^2, m^2 \right)}{m_W^2} \right]
\no & &
+ s_W^2 \left( 2 - s_W^2 \right)\, l \left( m_Z^2, m^2 \right),
\\
A_V &=& \frac{c_W^2}{4 s_W^2} \left[ k \left( m_Z^2, m^2, m^2 \right)
  - \frac{m^2}{m_Z^2}\, j \left( m_Z^2, m^2, m^2 \right) \right],
\\
A_W &=& \frac{1}{4 s_W^2} \left[ k \left( m_W^2, m^2, m^2 \right)
  - \frac{m^2}{m_W^2}\, j \left( m_W^2, m^2, m^2 \right) \right],
\\
A_X &=& \frac{c_W^2}{4}\ l \left( m_Z^2, m^2 \right),
\ea
\es
\bs
\label{eq_BO}
\ba
B_S = - B_U &=& 4 s_W^4\, l \left( m_Z^2, m^2 \right)
\\
B_V &=& \frac{s_W^2}{c_W^2} \left[ k \left( m_Z^2, m^2, m^2 \right)
  - \frac{m^2}{m_Z^2}\, j \left( m_Z^2, m^2, m^2 \right) \right],
\\
B_X &=& - s_W^2\, l \left( m_Z^2, m^2 \right),
\ea
\es
and $B_W = 0$.
All the coefficients $A_O$ and $B_O$ are increasing functions of $m$,
depicted in Fig.~\ref{1multp_ABo_vs_m}. 
Notice that,
in general,
the OPs $V$,
$W$,
and $X$ may be as important as $S$ and $U$.

%%%%% I ADDED
Using Eq.~\eqref{mflspdtk} one finds that
\be
B_S = - 4 s_W^4\, \frac{m_Z^2}{15 m^2}
+ \mathrm{O} \left( \frac{m_Z^4}{m^4} \right).
\label{bsapp}
\ee
Similarly,
using Eqs.~\eqref{6f},
\eqref{pipo1},
\eqref{pipo2},
\eqref{g000},
and \eqref{hatg000} one finds that
\be
A_S = - c_W^4\, \frac{m_Z^2}{15 m^2}
+ \mathrm{O} \left( \frac{m_Z^4}{m^4} \right).
\label{asapp}
\ee
Equations~\eqref{bsapp} and~\eqref{asapp} are excellent approximations
for the $B_S$ and $A_S$,
respectively,
depicted in Fig.~\ref{1multp_ABo_vs_m}.
%%%%%

\begin{figure}[ht]
\begin{center}
\includegraphics[width=0.5\textwidth]{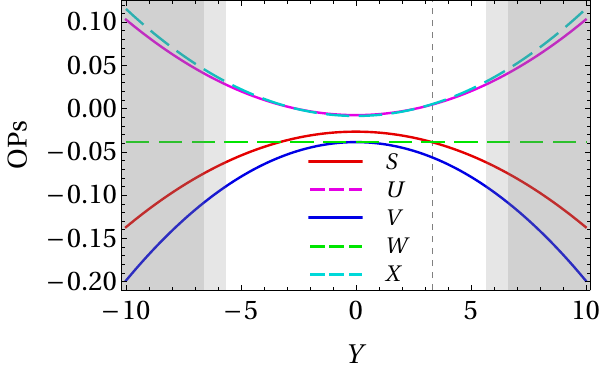}
\end{center}
%\vspace{-6mm}
\caption{The oblique parameters
  as functions of $Y$,
  while $n=5$ and $m=400$\,GeV are kept fixed.
  The dashed vertical line indicates
  the benchmark value $Y=3.3$.
  The light-gray area indicates that the corresponding OPs
  lead to a fit to the observables with $\chi^2 > 17$;
  for the dark-grey area one has $\chi^2 > 20$.}
\label{1multp_OPs_vs_Y}
\end{figure}

\begin{figure}[ht]
\begin{center}
\includegraphics[width=0.5\textwidth]{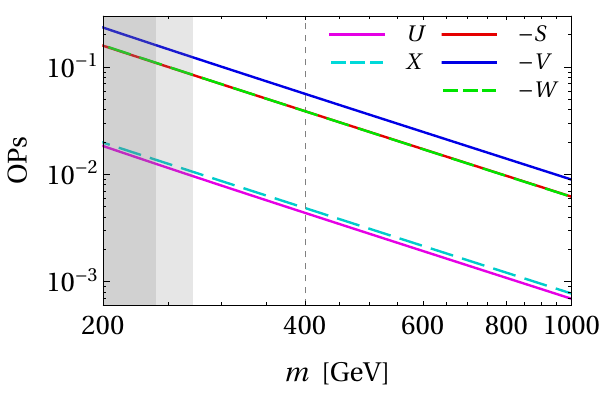}
\end{center}
%\vspace{-6mm}
\caption{The oblique parameters
  as functions of $m$,
  while $n=5$ and $Y=3.3$ are kept fixed.
  The dashed vertical line indicates
  the benchmark value $m=400$~GeV.
  The meaning of the gray-shadowed bands
  is the same as in Fig.~\ref{1multp_OPs_vs_Y}.
  For large $m$, all the oblique parameters are 
  approximately proportional to $m^{-2}$.  
%  The parameters obey following approximations: $S \approx W \approx -6700 \, m^{-2.01}$, $U \approx 800\, m^{-2.018}$, $V \approx -10500\, m^{-2.02}$ and $X \approx 840\, m^{-2.01}$.  
  }
\label{1multp_OPs_vs_m}
\end{figure}

\begin{figure}[ht]
\begin{center}
\includegraphics[width=1.0\textwidth]{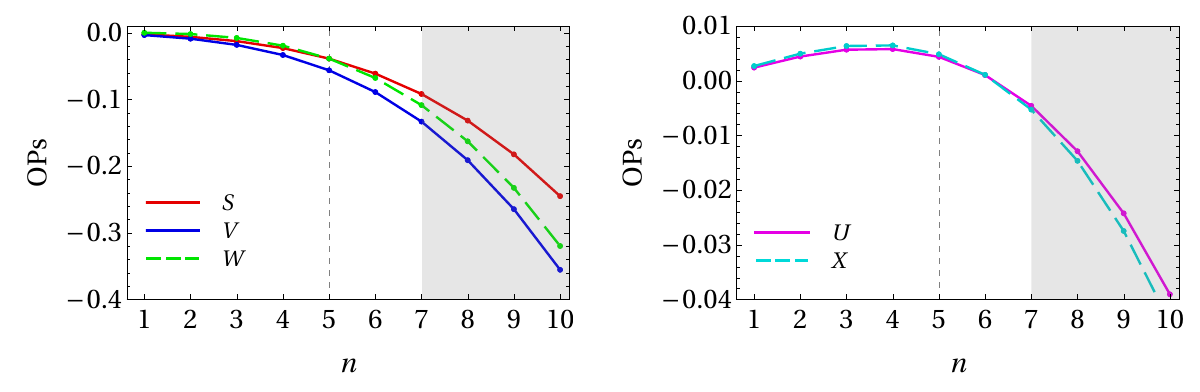}
\end{center}
%\vspace{-6mm}
\caption{The oblique parameters as functions of $n$,
  while $m=400$\,GeV and $Y=3.3$ are kept fixed.
  The dashed vertical line indicates the benchmark value $n=5$.
  The dark-gray area means that the OPs lead to $\chi^2 > 20$ fit.}
\label{1multp_OPs_vs_n}
\end{figure}

\begin{figure}[ht]
\begin{center}
\includegraphics[width=1.0\textwidth]{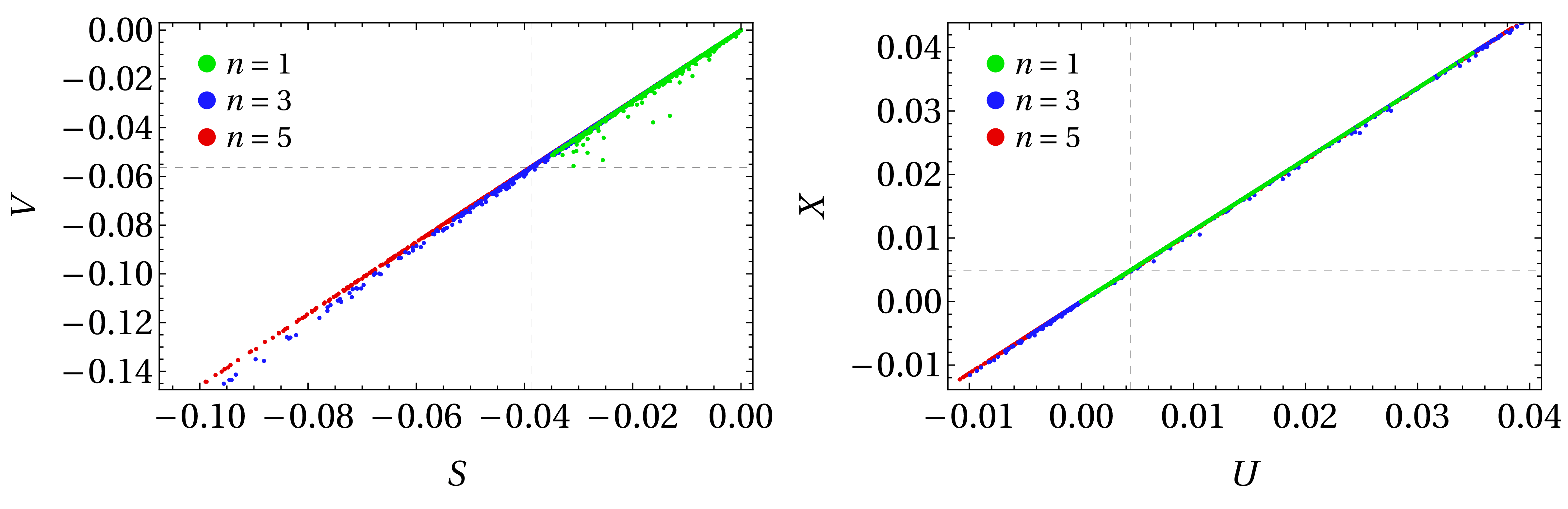}
\end{center}
%\vspace{-6mm}
\caption{Correlation plots between oblique parameters
  for different values of $n$.
  The parameters $S$ and $V$ are distributed according
  to $V \approx 1.47\, S$ (left panel),
  while the parameters $U$ and $X$
  obey $X \approx 1.12\, U$ (right panel).
  All points in the plots obey the restriction $\chi^2 \leq 20$.
  The dashed lines indicate the values of the oblique parameters
  at the benchmark point BP1.}
\label{1multp_OPs_vs_Ops}
\end{figure}

This New Physics Model gives a fit of the OPs which is just a little worse
than letting the OPs vary freely.
Indeed,
by setting $V = W = X = 0$ and allowing $S$,
$T$,
and $U$ to vary freely\footnote{Our best fit was obtained for
$S = -1.2 \times 10^{-2}$,
$T = 2.8 \times 10^{-2}$,
and $U = 2.0 \times 10^{-3}$.}
we were able to accomplish a fit
of all the relevant electroweak observables\footnote{We have used
the following twenty observables,
taken from Ref.~\cite{PDG2022}:
$R_\ell$,
$R_b$,
$R_c$,
$A_\ell$,
$A_b$,
$A_c$,
$A_{FB}^{\left( 0, \ell \right)}$,
$A_{FB}^{\left( 0, b \right)}$,
$A_{FB}^{\left( 0, c \right)}$,
$g_V^{\nu e}$,
$g_A^{\nu e}$,
$\bar{s}_\ell^2$ (three different values),
$m_W$,
$\Gamma_W$,
$\Gamma_Z$,
$\sigma_\mathrm{had}$,
$Q_W \! \left( \mathrm{Cs} \right)$,
and $Q_W \! \left( \mathrm{Tl} \right)$.}
with $\chi^2 = 14.201$;
while in our NPM with $m = 400$\,GeV,
$n = 5$,
and $Y = 3.3$ we achieve $\chi^2 = 14.894$,
which is not much worse\footnote{We perform a fit
by defining $\chi^2 = R C^{-1} R^T$,
where $R$ is the row-vector of the residuals of the observables
and $C$ is the covariance matrix,
which is evaluated according to the correlations
among the observables~\cite{PDG2022,ALEPH,Tenchini}.}.
We use the above values of $m$,
$n$,
and $Y$ as our first benchmark point (BP1).
Then,
\begin{itemize}
\item Keeping both $n$ and $m$ fixed at their BP1 values,
  we let $Y$ vary and observe
  the variation of the OPs displayed in Fig.~\ref{1multp_OPs_vs_Y}.
\item Keeping both $n$ and $Y$ fixed at their BP1 values,
  we let $m$ vary and observe
  the variation of the OPs displayed in Fig.~\ref{1multp_OPs_vs_m}.
\item Keeping both $Y$ and $m$ fixed at their BP1 values,
  we let $n$ vary and observe
  the variation of the OPs displayed in Fig.~\ref{1multp_OPs_vs_n}.
\end{itemize}
We also observe that there are
approximate linear correlations between the parameters $S$ and $V$,
and between the parameters $U$ and $X$,
displayed in Fig.~\ref{1multp_OPs_vs_Ops}.

A more detailed description of the numerical analyses
is given in subsection~\ref{sec:numerics}.

\section{Two vector-like multiplets}
\label{sec:two}

Since the formalism in section~\ref{sec:mixing} may look a bit abstract,
we give in this section the practical calculation
of the mixing matrices in a specific NPM with \emph{vector-like}
(in order to avoid anomalies)
fermions.\footnote{The NPM that we deal with in this section
has been recently suggested in Ref.~\cite{nortier}.}
In our model all the fermion masses are justified
either through bare mass terms
or through $SU(2) \times U(1)$-invariant Yukawa couplings
to the Higgs doublet of the SM;
therefore,
the oblique parameter $T$ has no reason to feature an UV divergence and,
indeed,
it converges.

\subsection{Description of the model}

In the NPM that we suggest there are,
besides all the fermion multiplets and scalar multiplets of the SM,
the following multiplets of fermions:\footnote{For the sake of simplicity,
we assume all the new fermions to be color singlets.}
\begin{itemize}
\item One multiplet $A_L$ of left-handed fermions
  with isospin $J$ and hypercharge $Y$.
\item One multiplet $B_L$ of left-handed fermions
  with isospin $J+1/2$ and hypercharge $Y+1/2$.
\item Two multiplets $A_R$ and $B_R$ of right-handed fermions
  with the same quantum numbers as those of $A_L$ and $B_L$,
  respectively.
\end{itemize}
We define $n \equiv 2 J + 1 $.
We write the multiplets of additional fermions as
\be
\label{um}
A_E = \left( \begin{array}{c} a_{1,E} \\ a_{2,E} \\
  \vdots \\ a_{n,E} \end{array} \right),
\quad
B_E = \left( \begin{array}{c} b_{0,E} \\ b_{1,E} \\ b_{2,E} \\
  \vdots \\ b_{n,E} \end{array} \right).
\ee
%
%Let $T_3$ be the third component of isospin;
%then,
%
%\bs
%\ba
%T_3\, a_{k,E} &=& \frac{n + 1 - 2 k}{2}\, a_{k,E},
%\\
%T_3\, b_{j,E} &=& \frac{n - 2 j}{2}\, b_{j,E}.
%\ea
%\es
%
There are bare-mass terms given by
\be
\label{jf94994}
\mathcal{L}_\mathrm{bare} =
- m_A \sum_{k=1}^n\, \overline{a_{k,R}}\ a_{k,L}
- m_B \sum_{j=0}^n\, \overline{b_{j,R}}\ b_{j,L}
+ \mathrm{H.c.}
\ee
The quantum numbers of the new fermion multiplets were chosen
in such a way that they have
$SU(2) \times U(1)$-invariant Yukawa couplings to the Higgs doublet of the SM
$\left( \varphi^+, \ \varphi^0 \right)^T$,
which has isospin and hypercharge $1/2$.
It is easy to convince oneself that the Yukawa couplings
of $\varphi^0$ to the new fermions are given by
\be
\mathcal{L}_\mathrm{Yukawa} = \cdots - \varphi^0\ \sum_{k=1}^n \sqrt{k}\, \left(
y_R\ \overline{b_{k,R}}\ a_{k,L} + y_L\ \overline{b_{k,L}}\ a_{k,R} \right)
+ \mathrm{H.c.},
\label{hvufid}
\ee
with Yukawa coupling constants $y_R$ and $y_L$.
%In Eq.~\eqref{hvufid} we have chosen not to display
%the Yukawa couplings of $\varphi^+$.
Since the largest Yukawa couplings are $y_E \sqrt{n}$,
we assume that
\be
\label{jbicco}
\left| y_R \right|, \left| y_L \right| < \frac{4 \pi}{\sqrt{n}}
\ee
in order to respect unitarity.\footnote{Stronger
unitarity constraints may exist,
arising for instance from the scattering of fermions into gauge-boson pairs;
see Ref.~\cite{barducci} and,
for the case of large \emph{scalar} multiplets,
see Ref.~\cite{logan}.}

In Eq.~\eqref{hvufid},
note that $b_{0,L}$ and $b_{0,R}$ have no Yukawa couplings to $\varphi^0$.
Together they form a Dirac fermion with electric charge
$\left. \left( n+1 \right) \right/ 2 + Y$.
Its mass term is
\be
- m_B\ \overline{b_{0,R}}\ b_{0,L} + \mathrm{H.c.}
\ee
For $k = 1, \ldots, n$,
there are \emph{two} Dirac fermions with electric
charge $\left. \left( n+1 \right) \right/ 2 + Y - k$.
According to Eqs.~\eqref{jf94994} and~\eqref{hvufid},
their mass terms are given by
\be
\label{kgpdrc}
- \left( \begin{array}{cc} \overline{b_{k,R}}, & \overline{a_{k,R}}
\end{array} \right)
\left( \begin{array}{cc} m_B & \sqrt{k}\, m_C \\
  \sqrt{k}\, m_D & m_A \end{array} \right)
\left( \begin{array}{c} b_{k,L} \\ a_{k,L} \end{array} \right).
\ee
In Eq.~\eqref{kgpdrc},
$m_C \equiv y_R v$ and $m_D \equiv y_L^\ast v^\ast$,
where $v$ is the VEV of $\varphi^0$,
with $\left| v \right| \approx 174$\,GeV.
According to Eq.~\eqref{jbicco},
\bs
\label{jguc}
\ba
\left| m_C \right| &<& \frac{4 \pi \left| v \right|}{\sqrt{n}},
\\
\left| m_D \right| &<& \frac{4 \pi \left| v \right|}{\sqrt{n}},
\ea
\es
while $\left| m_A \right|$ and $\left| m_B \right|$
may be as large as one wishes.
Our NPM has six real free parameters:
$Y$,
$\arg{\left( m_A m_B m_C^\ast m_D^\ast \right)}$,
and
\bs
\label{utyreq}
\ba
a &\equiv& \left| m_A \right|^2, \\
b &\equiv& \left| m_B \right|^2, \\
c &\equiv& \left| m_C \right|^2, \\
d &\equiv& \left| m_D \right|^2.
\ea
\es
Additionally there is $n$,
which is an integer.

For $k = 1, \ldots, n$,
we diagonalize the mass matrix in Eq.~\eqref{kgpdrc} by making
\be
\label{pdpdp}
\left( \begin{array}{c} b_{k,E} \\ a_{k,E} \end{array} \right)
=
\mathcal{U}_{k,E}
\left( \begin{array}{c} f_{k,E} \\ g_{k,E} \end{array} \right),
\ee
where the $2 \times 2$ matrices $\mathcal{U}_{k,E}$
are unitary
and the physical fermions $f_k$ and $g_k$ have masses $m_{f,k}$ and $m_{g,k}$,
respectively.
We define
\be
\label{mk}
M_k = \left( \begin{array}{cc} m_{f,k} & 0 \\ 0 & m_{g,k} \end{array} \right).
\ee
The matrices $M_k$ are diagonal and real.
The bi-diagonalization condition is
\be
\label{jbigfifi}
\left( \begin{array}{cc} m_B & \sqrt{k}\, m_C \\
  \sqrt{k}\, m_D & m_A \end{array} \right)
= \mathcal{U}_{k,R}\, M_k\, \mathcal{U}_{k,L}^\dagger.
\ee

It is convenient to write
\be
\label{cmbkfpd}
\mathcal{U}_{k,E} = \left( \begin{array}{c} X_{k,E} \\ Y_{k,E}
\end{array} \right),
\ee
where $X_{k,E}$ and $Y_{k,E}$ are $1 \times 2$ matrices.
Thus,
from Eq.~\eqref{pdpdp},
\bs
\label{89ffc}
\ba
b_{k,E} &=&
X_{k,E} \left( \begin{array}{c} f_{k,E} \\ g_{k,E} \end{array} \right),
\\
a_{k,E} &=&
Y_{k,E} \left( \begin{array}{c} f_{k,E} \\ g_{k,E} \end{array} \right),
\\
\left( \begin{array}{c} f_{k,E} \\ g_{k,E} \end{array} \right)
%= \mathcal{U}_{k,E}^\dagger
%\left( \begin{array}{c} b_{k,E} \\ a_{k,E} \end{array} \right)
&=& X_{k,E}^\dagger b_{k,E} + Y_{k,E}^\dagger a_{k,E}.
\ea
\es
The unitarity of $\mathcal{U}_{k,E}$ implies
\bs
\label{unita}
\ba
X_{k,E}^\ast X_{k,E}^T &=& 1, \label{66a} \\
Y_{k,E}^\ast Y_{k,E}^T &=& 1, \label{66b} \\
X_{k,E}^\ast Y_{k,E}^T = Y_{k,E}^\ast X_{k,E}^T &=& 0, \label{66c} \\
X_{k,E}^T X_{k,E}^\ast + Y_{k,E}^T Y_{k,E}^\ast &=& \mathbf{1}_2,
\label{14e}
\ea
\es
where $\mathbf{1}_2$ is the $2 \times 2$ unit matrix.
From Eqs.~\eqref{jbigfifi} and~\eqref{cmbkfpd},
\bs
\label{7849}
\ba
m_B &=& X_{k,R}\, M_k\, X_{k,L}^\dagger, \\
m_A &=& Y_{k,R}\, M_k\, Y_{k,L}^\dagger, \\
\sqrt{k}\, m_C &=& X_{k,R}\, M_k\, Y_{k,L}^\dagger, \\
\sqrt{k}\, m_D &=& Y_{k,R}\, M_k\, X_{k,L}^\dagger.
\ea
\es
Utilizing Eq.~\eqref{14e} and remembering that $M_k = M_k^\dagger$,
one may derive from Eqs.~\eqref{7849} that
\bs
\label{71}
\ba
b + k c &=& X_{k,R}\, M_k^2\, X_{k,R}^\dagger, \label{68a} \\
a + k d &=& Y_{k,R}\, M_k^2\, Y_{k,R}^\dagger, \label{68b} \\
b + k d &=& X_{k,L}\, M_k^2\, X_{k,L}^\dagger, \label{68d} \\
a + k c &=& Y_{k,L}\, M_k^2\, Y_{k,L}^\dagger, \label{68e}
\ea
\es
where $a$,
$b$,
$c$,
and $d$ have been defined in Eqs.~\eqref{utyreq}.

\subsection{The mixing matrices}

We now apply our formalism to the model described in the previous subsection.
Firstly,
we put together all the physical fermions of each chirality
in column vectors
\be
V_E = \left( \begin{array}{c} b_{0,E} \\ f_{1, E} \\ g_{1,E} \\
  \vdots \\ f_{n,E} \\ g_{n,E}
\end{array} \right),
\ee
taking care \emph{to order the fermions by their decreasing electric charges}.
Indeed,
the (diagonal) electric-charge matrix
for the $2 n + 1 $ physical fermions in $V_E$ is
\be
\label{15}
Q = \frac{1}{2} \left( \begin{array}{cccccc}
  n + 1 & 0_{1 \times 2} & 0_{1 \times 2} & \cdots & 0_{1 \times 2} \\
  0_{2 \times 1} & \left( n - 1  \right) \times \mathbf{1}_2 & 0_{2 \times 2}
  & \cdots & 0_{2 \times 2} \\
  0_{2 \times 1} & 0_{2 \times 2} & \left( n - 3  \right) \times \mathbf{1}_2
  & \cdots & 0_{2 \times 2} \\
  \vdots & \vdots & \vdots & \ddots & \vdots \\
  0_{2 \times 1} & 0_{2 \times 2} & 0_{2 \times 2} & \cdots &
  \left( 1 - n \right) \times \mathbf{1}_2
\end{array} \right) + Y \times \mathbf{1}_{2 n + 1}.
\ee
The (diagonal) mass matrix of the physical fermions in $V_E$ is
\be
\label{16}
M = \left( \begin{array}{ccccc}
  m_B & 0_{1 \times 2} & 0_{1 \times 2} & \ldots & 0_{1 \times 2} \\
  0_{2 \times 1} & M_1 & 0_{2 \times 2} & \ldots & 0_{2 \times 2} \\
  0_{2 \times 1} & 0_{2 \times 2} & M_2 & \ldots & 0_{2 \times 2} \\
  \vdots & \vdots & \vdots & \ddots & \vdots \\
  0_{2 \times 1} & 0_{2 \times 2} & 0_{2 \times 2} & \cdots & M_n
\end{array} \right),
\ee
where the matrices $M_k$ have been defined in Eq.~\eqref{mk}.

We define the matrices $M_E \left/ \sqrt{2} \right.$,
which represent the action of the $SU(2)$ operator $T_+$
on the fermions of $V_E$,
through
\be
V_E^T\, \frac{M_E}{\sqrt{2}} = \left( \begin{array}{cccccc}
  \left( T_+ b_{0,E} \right), & \left( T_+ f_{1,E} \right), &
  \left( T_+ g_{1,E} \right), & \ldots, &
  \left( T_+ f_{n,E} \right), & \left( T_+ g_{n,E} \right)
\end{array} \right).
\ee
Obviously,
$T_+ b_{0,E} = 0$.
Now,
utilizing Eqs.~\eqref{89ffc},
\bs
\label{uvifof3}
\ba
\sqrt{2}\, T_+ \left( \begin{array}{c} f_{1,E} \\ g_{1,E} \end{array} \right)
&=& \sqrt{2}\, T_+ \left( X_{1,E}^\dagger b_{1,E} + Y_{1,E}^\dagger a_{1,E} \right)
\\ &=& X_{1,E}^\dagger\, \sqrt{n}\ b_{0,E},
\ea
\es
and,
for $m = 1, \ldots, n-1$,
\bs
\ba
\sqrt{2}\, T_+
\left( \begin{array}{c} f_{m+1,E} \\ g_{m+1,E} \end{array} \right)
&=& \sqrt{2}\, T_+
\left( X_{m+1,E}^\dagger b_{m+1,E} + Y_{m+1,E}^\dagger a_{m+1,E} \right)
\\ &=& \sqrt{n - m}\, \left( X_{m+1,E}^\dagger \sqrt{m+1}\, b_{m,E}
+ Y_{m+1,E}^\dagger \sqrt{m}\, a_{m,E} \right)
\\ &=& \sqrt{n - m}
\left( \sqrt{m+1}\ X_{m+1,E}^\dagger X_{m,E}
%\right. \no & & \left.
+ \sqrt{m}\ Y_{m+1,E}^\dagger Y_{m,E} \right)
\left( \begin{array}{c} f_{m,E} \\ g_{m,E} \end{array} \right).
\hspace*{8mm}
\ea
\es
Therefore,
\be
\label{meme0}
M_E = \left( \begin{array}{cccccc}
  0 & M_{E,0} & 0_{1 \times 2} & 0_{1 \times 2} & \cdots & 0_{1 \times 2} \\
  0_{2 \times 1} & 0_{2 \times 2} & M_{E,1} & 0_{2 \times 2} & \cdots & 0_{2 \times 2} \\
  0_{2 \times 1} & 0_{2 \times 2} & 0_{2 \times 2} & M_{E,2} & \cdots & 0_{2 \times 2} \\
  \vdots & \vdots & \vdots & \vdots & \ddots & \vdots\\
  0_{2 \times 1} & 0_{2 \times 2} & 0_{2 \times 2} & 0_{2 \times 2} &
  \cdots & M_{E,n-1} \\
  0_{2 \times 1} & 0_{2 \times 2} & 0_{2 \times 2} & 0_{2 \times 2} &
  \cdots & 0_{2 \times 2}
\end{array} \right),
\ee
where,
from Eq.~\eqref{uvifof3},
\be
\label{Ms}
M_{E,0} = \sqrt{n}\, X_{1,E}^\ast
\ee
is a $1 \times 2$ matrix,
and the
\be
M_{E,m} = \sqrt{n - m} \left(
\sqrt{m+1}\ X_{m,E}^T X_{m+1,E}^\ast
+ \sqrt{m}\ Y_{m,E}^T Y_{m+1,E}^\ast \right)
\qquad (m = 1, \ldots, n-1)
\label{ibugod}
\ee
are $2 \times 2$ matrices.
Notice that $M_E$ in Eq.~\eqref{meme0},
just like $Q$ in Eq.~\eqref{15} and $M$ in Eq.~\eqref{16},
is a $\left( 2 n + 1 \right) \times \left( 2 n + 1 \right)$ matrix,
because there are $2 n + 1$ (new) Dirac fermions in our NPM.

For $k = 1, \ldots, n$ we define the $2 \times 2$ Hermitian,
idempotent matrices
\bs
\ba
H_{k,E} &\equiv& X_{k,E}^T X_{k,E}^\ast,
\\
\left( H_{k,E} \right)^2 &=& H_{k,E}. \label{cbvofpd}
\ea
\es
(Equation~\eqref{cbvofpd} follows from Eq.~\eqref{66a}.)
It is then easy to see that
\be
M_E M_E^\dagger = \left( \begin{array}{cccccc}
  M_{E,0} M_{E,0}^\dagger & 0_{1 \times 2} & 0_{1 \times 2} &
  \cdots & 0_{1 \times 2} & 0_{1 \times 2} \\
  0_{2 \times 1} & M_{E,1} M_{E,1}^\dagger & 0_{2 \times 2} &
  \cdots & 0_{2 \times 2} & 0_{2 \times 2} \\
  0_{2 \times 1} & 0_{2 \times 2} & M_{E,2} M_{E,2}^\dagger &
  \cdots & 0_{2 \times 2} & 0_{2 \times 2} \\
  \vdots & \vdots & \vdots & \ddots & \vdots & \vdots \\
  0_{2 \times 1} & 0_{2 \times 2} & 0_{2 \times 2} & \cdots &
  M_{E,n-1} M_{E,n-1}^\dagger & 0_{2 \times 2} \\
  0_{2 \times 1} & 0_{2 \times 2} & 0_{2 \times 2} & \cdots &
  0_{2 \times 2} & 0_{2 \times 2}
\end{array} \right)
\ee
has,
because of Eqs.~\eqref{unita},
\bs
\ba
M_{E,0} M_{E,0}^\dagger &=& n, \\
M_{E,m} M_{E,m}^\dagger &=& \left( n - m \right) \left( m \times \mathbf{1}_2
+ H_{m,E} \right) \qquad (m = 1, \ldots, n-1);
\ea
\es
while
\be
M_E^\dagger M_E = \left( \begin{array}{cccccc}
  0 & 0_{1 \times 2} & 0_{1 \times 2} &
  \cdots & 0_{1 \times 2} & 0_{1 \times 2} \\
  0_{2 \times 1} & M_{E,0}^\dagger M_{E,0} & 0_{2 \times 2} &
  \cdots & 0_{2 \times 2} & 0_{2 \times 2} \\
  0_{2 \times 1} & 0_{2 \times 2} & M_{E,1}^\dagger M_{E,1} &
  \cdots & 0_{2 \times 2} & 0_{2 \times 2} \\
  \vdots & \vdots & \vdots & \ddots & \vdots & \vdots \\
  0_{2 \times 1} & 0_{2 \times 2} & 0_{2 \times 2} & \cdots &
  M_{E,n-2}^\dagger M_{E,n-2} & 0_{2 \times 2} \\
  0_{2 \times 1} & 0_{2 \times 2} & 0_{2 \times 2} & \cdots &
  0_{2 \times 2} & M_{E,n-1}^\dagger M_{E,n-1}
\end{array} \right)
\ee
has
\bs
\ba
M_{E,0}^\dagger M_{E,0} &=& n \times H_{1,E}, \\
M_{E,m}^\dagger M_{E,m} &=& \left( n - m \right) \left( m \times \mathbf{1}_2
+ H_{m+1,E} \right) \qquad (m = 1, \ldots, n-1).
\ea
\es
Then,
according to the definition~\eqref{he},
\be
\label{he0}
H_E = M_E M_E^\dagger - M_E^\dagger M_E = \left( \begin{array}{ccccc}
  H_{E,0} & 0_{1 \times 2} & 0_{1 \times 2} & \cdots & 0_{1 \times 2} \\
  0_{2 \times 1} & H_{E,1} & 0_{2 \times 2} & \cdots & 0_{2 \times 2} \\
  0_{2 \times 1} & 0_{2 \times 2} & H_{E,2} & \cdots & 0_{2 \times 2} \\
  \vdots & \vdots & \vdots & \ddots & \cdots \\
  0_{2 \times 1} & 0_{2 \times 2} & 0_{2 \times 2} & \cdots & H_{E,n}
\end{array} \right),
\ee
where
\bs
\ba
\label{he1}
H_{E,0} &=& n,
\\
\label{he2}
H_{E,k} &=& \left( n + 1 - 2 k \right) \times \mathbf{1}_2
- H_{k,E} \qquad (k = 1, \ldots, n).
\ea
\es
Finally,
the $\left( 2 n + 1 \right) \times \left( 2 n + 1 \right)$ Hermitian matrices
$F_E$ are given by Eqs.~\eqref{F},
\eqref{15},
and~\eqref{he0}.

\subsection{The finiteness of $T$}

For completeness,
in this subsection we explicitly demonstrate
that Eq.~\eqref{fjvfio} holds in our NPM and that,
therefore,
the oblique parameter $T$ is finite in it.

One may define
\be
x_E \equiv \left\{ \begin{array}{ll}
    c & \Leftarrow\ E = R, \\
    d & \Leftarrow\ E = L.
\end{array} \right.
\ee
Then,
from Eqs.~\eqref{68a} and~\eqref{68d},
\be
\mathrm{tr} \left( H_{k,E} M_k^2 \right)
= b + k x_E, \qquad (k = 1, \ldots n).
\ee
Also,
Eqs.~\eqref{68a},
\eqref{68b},
and~\eqref{14e} imply
\be
\mathrm{tr} \left( M_k^2 \right) = a + b
+ k \left( c + d \right) \qquad (k = 1, \ldots, n ).
\ee

It is then easy to derive that
\bs
\ba
\mathrm{tr} \left( M_E M_E^\dagger M^2 \right) &=& n b
+ \sum_{m=1}^{n-1} \left( n - m \right)
\left\{ m \left[ a + b + m  \left( c + d \right) \right] + b + m x_E \right\},
\\
\mathrm{tr} \left( M_E^\dagger M_E M^2 \right) &=& n \left( b + x_E \right)
+ \sum_{m=1}^{n-1} \left( n - m \right)
\no & & \times
\left\{ m \left[ a + b + \left( m + 1 \right) \left( c + d \right) \right]
+ b + \left( m + 1 \right) x_E \right\},
\\
\mathrm{tr} \left( H_E^2 M^2 \right) &=& n^2 b
+ \sum_{k=1}^n \left\{ \left( n + 1 - 2 k \right)^2
\left[ a + b + k \left( c + d \right) \right]
\right. \no & & \left.
+ \left( 4 k - 2 n - 1 \right) \left( b + k x_E \right) \right\}.
\ea
\es
Performing the sums over $m$ by using
\bs
\ba
\sum_{m=1}^{n-1}\, 1 &=& n - 1, \\
\sum_{m=1}^{n-1}\, m &=& \left( n - 1 \right)\, \frac{n}{2}, \\
\sum_{m=1}^{n-1}\, m^2 &=& \left( n - 1 \right)\,
\frac{n \left( 2 n - 1 \right)}{6}, \\
\sum_{m=1}^{n-1}\, m^3 &=& \left( n - 1 \right)^2\,
\frac{n^2}{4},
\ea
\es
one finds that
\be
\mathrm{tr}
\left[ \left( M_E M_E^\dagger + M_E^\dagger M_E - H_E^2 \right) M^2 \right] = 0
\ee
for both $E = L$ and $E = R$.
Therefore,
each of the two lines~\eqref{tra} and~\eqref{trb} separately vanishes.

One also finds that
\bs
\ba
\mathrm{tr} \left( H_L M H_R M \right) &=&
b H_{L,0} H_{R,0} + \sum_{k=1}^n \mathrm{tr} \left( H_{L,k} M_k H_{R,k} M_k \right)
\\ &=& b n^2 + \sum_{k=1}^n \left\{ \left( n + 1 - 2 k \right)^2
  \mathrm{tr} \left( M_k^2 \right)
  \right. \no & & \left.
  - \left( n + 1 - 2 k \right) \mathrm{tr}
  \left[ \left( H_{k,L} + H_{k,R} \right) M_k^2 \right]
  + \mathrm{tr} \left( H_{k,L} M_k H_{k,R} M_k \right) \right\}
  \hspace*{7mm}
\\ &=&
b n^2
+ \sum_{k=1}^n \left\{ \left( n + 1 - 2 k \right)^2
\left[ a + b + k  \left( c + d \right) \right]
\right. \no & & \left.
- \left( n + 1 - 2 k \right) \left[ 2 b + k  \left( c + d \right) \right]
+ b \right\},
\\
\mathrm{tr} \left( M_L M M_R^\dagger M \right) &=&
m_B \left( M_{L,0} M_1 M_{R,0}^\dagger \right)
+ \sum_{m=1}^{n-1} \mathrm{tr} \left( M_{L,m} M_{m+1} M_{R,m}^\dagger M_m \right)
\no &=&
m_B n \left( X_{1,L}^\ast M_1 X_{1,R}^T \right)
\no & &
+ \sum_{m=1}^{n-1} \left( n - m \right) \mathrm{tr} \left[
  \vphantom{\sqrt{m \left( m + 1 \right)}}
  \left( m + 1 \right) X_{m,L}^T X_{m+1,L}^\ast M_{m+1} X_{m+1,R}^T X_{m,R}^\ast M_m
  \right. \no & &
  + m Y_{m,L}^T Y_{m+1,L}^\ast M_{m+1} Y_{m+1,R}^T Y_{m,R}^\ast M_m
  \no & &
  + \sqrt{m \left( m + 1 \right)}\, X_{m,L}^T X_{m+1,L}^\ast M_{m+1}
  Y_{m+1,R}^T Y_{m,R}^\ast M_m
  \no & & \left.
  + \sqrt{m \left( m + 1 \right)}\, Y_{m,L}^T Y_{m+1,L}^\ast M_{m+1}
  X_{m+1,R}^T X_{m,R}^\ast M_m \right]
\no &=&
n b + \sum_{m=1}^{n-1} \left( n - m \right)
\left[ \left( m + 1 \right) b + m a
  + m \left( m + 1 \right) \left( c + d \right) \right].
\ea
\es
Therefore,
once again performing the sums over $m$,
\be
\mathrm{tr} \left( 2\, M_L M M_R^\dagger M - H_L M H_R M \right) = 0
\ee
and line~\eqref{jvifo0} is zero.
Thus,
in our model
\emph{each of the three lines of Eq.~\eqref{fjvfio} is separately zero}.

\subsection{Numerical results}
\label{sec:numerics}

%In this NPM there are six real free parameters,
%plus the integer $n$,
%to fit the OPs. 
%This allows us to achieve an even slightly smaller $\chi^2$
%than the one given by our $STU$ electroweak fit.
Our benchmark point 2 (BP2)
has $\left| m_A \right| = \left| m_B \right| = 2000$\,GeV,
$\left| m_C \right| = \left| m_D \right| = 100$\,GeV,
$\arg{\left( m_A^\ast m_B^\ast m_C m_D \right)} = 1.5$,
$Y=3.3$,
and $n=5$.
This yields a fit to the twenty electroweak observables
with $\chi^2 = 14.214$,
which is comparable to our best fit with null $V$,
$W$,
and $X$.

\begin{figure}[ht]
\begin{center}
\includegraphics[width=0.5\textwidth]{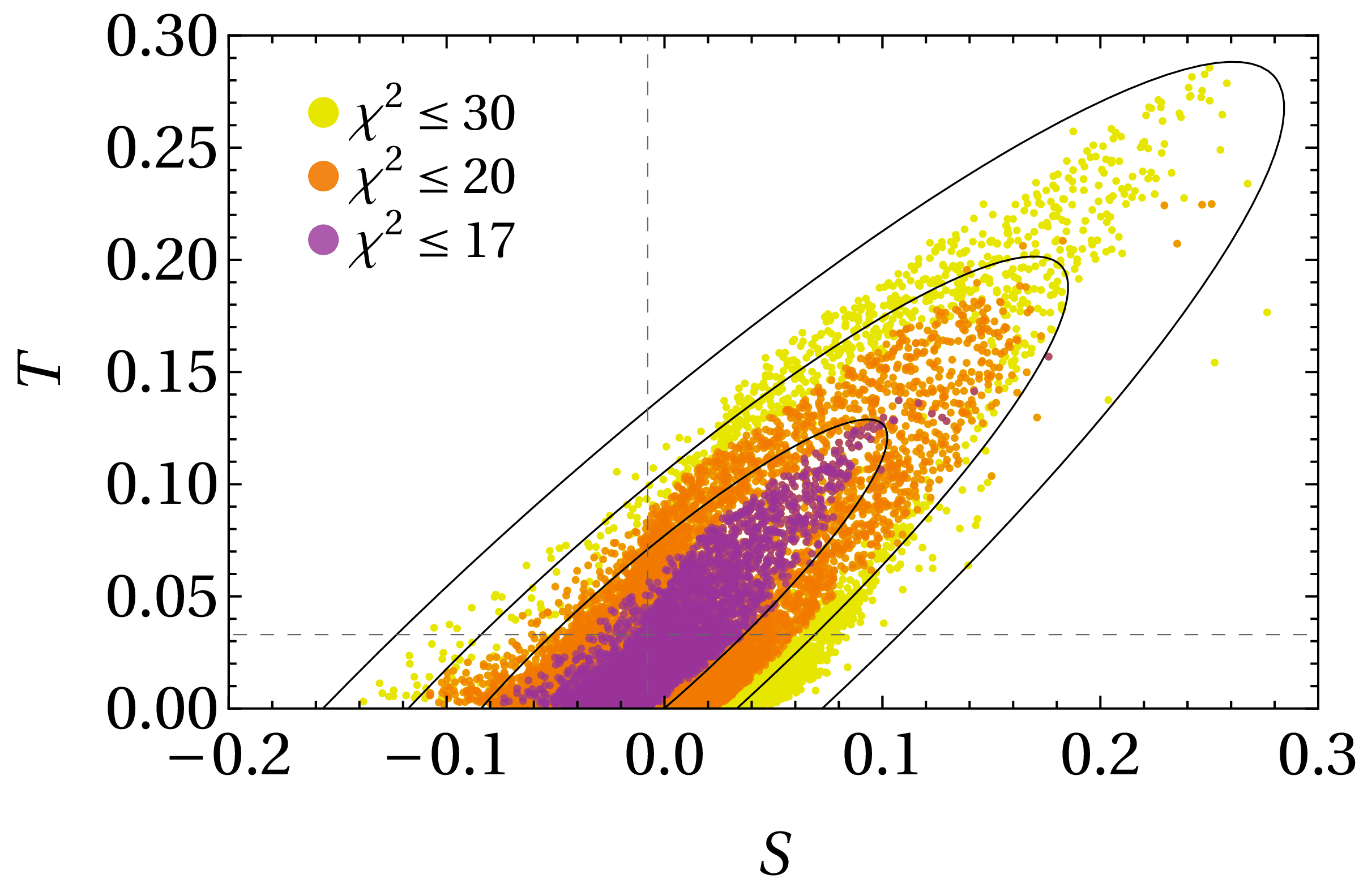}
\end{center}
%\vspace{-6mm}
\caption{The correlation between the oblique parameters $S$ and $T$
  in our New Physics Model,
  for $n=5$.
  The black ellipses correspond to the $1\sigma$,
  $2\sigma$,
  and $3\sigma$~(2dof) allowed regions in the $ST$ plane for
  a fit with $U = V = W = X = 0$ and completely free $S$ and $T$.
  The dashed lines indicate the values of the oblique parameters
  at the benchmark point BP2.}
\label{2multp_ellipses}
\end{figure}

In order to explore the entire parameter space,
we consider various integer values of $n$,
we let the masses vary from 50\,GeV to 3000\,GeV
but subject to the constraints~\eqref{jguc},
we let $\arg{\left( m_A^\ast m_B^\ast m_C m_D \right)}$ vary from $0$ to $2 \pi$,
and we let $Y$ go from $-10$ to $+10$.
We keep only the points that have $\chi^2$ smaller than a certain number,
which may be either 30,
20,
or 17.\footnote{The pull of observable $O$ is defined as
$\left. \left( O_\mathrm{fit} - O_\mathrm{measured} \right) \right/
\delta^\pm_\mathrm{measured}$,
where $O_\mathrm{measured}$ is the central value
and $\delta^\pm_\mathrm{measured}$ is the error in the measurement of $O$.
%The function $\chi^2$ is the sum of the squares of the pulls
%of all twenty observables.
In practice,
most pulls are always very small
and only very few observables have large pulls.
As a consequence,
points with $\chi^2 < 30$ have all the pulls between $-3$ and $+3$;
points with $\chi^2 < 20$ have pulls ranging from $-2$ to $+2$,
except for the observables $A_{FB}^{\left( 0, b \right)}$ and $A_\ell$;
and points with $\chi^2 < 17$ have pulls between $-1$ and $+1$,
with the additional exceptions of $R_\ell$
and $Q_W\!\left(\mathrm{Cs}\right)$.}
This differentiation of the points according to their $\chi^2$
coincides well with the correlation between the $S$ and $T$ parameters
in the electroweak fit,
displayed in Fig.~\ref{2multp_ellipses}.
This figure also shows that our NPM can only produce
\emph{positive} values for the parameter $T$.

\begin{figure}[ht]
\begin{center}
\includegraphics[width=0.5\textwidth]{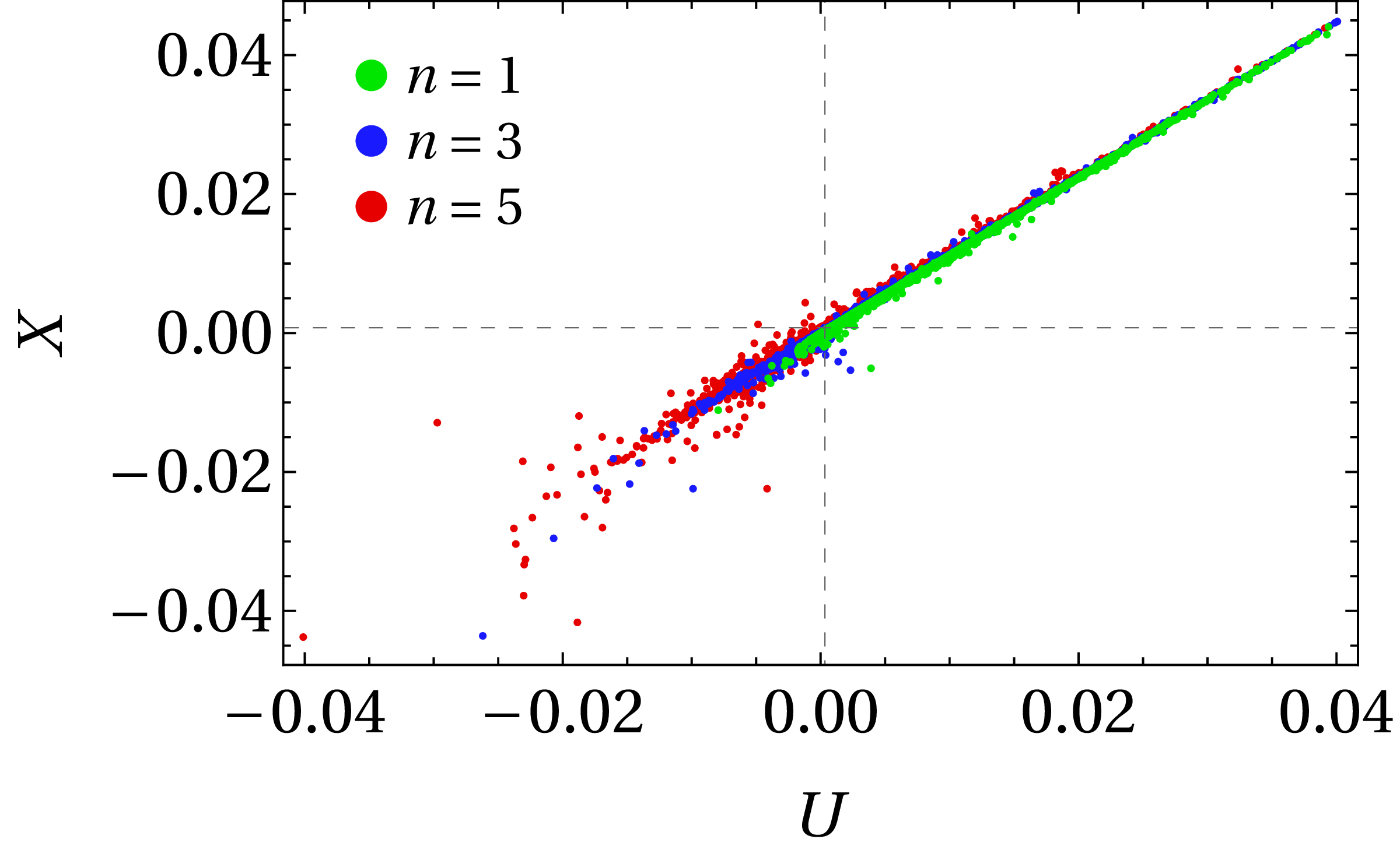}
\end{center}
%\vspace{-6mm}
\caption{The correlation between the OPs $U$ and $X$,
  for different values of $n$.
  All the points in this plot have $\chi^2 \leq 20$.
  The dashed lines indicate the values of the OPs in the BP2.}
\label{2multp_X_vs_U}
\end{figure}

In our NPM there is the approximate linear correlation
between the oblique parameters $U$ and $X$
displayed in Fig.~\ref{2multp_X_vs_U}.
The distribution of parameters is very similar
to the one observed for the NPM of section~\ref{sec:one},
\ie\ here too one has $X \approx 1.12\, U$.

\begin{figure}[ht]
\begin{center}
\includegraphics[width=0.5\textwidth]{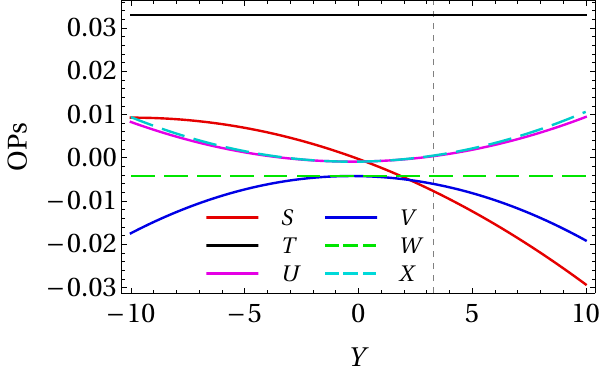}
\end{center}
%\vspace{-6mm}
\caption{The OPs as functions of $Y$.
  The dashed vertical line indicates the BP2 value $Y=3.3$.}
\label{2multp_Ops_vs_Y}
\end{figure}

In Fig.~\ref{2multp_Ops_vs_Y}
the variation of the OPs with $Y$ is displayed;
all the other parameters of the model are kept fixed at their BP2 values.
As ever,
the OPs $W$ and $T$ are constant because Eqs.~\eqref{W} and~\eqref{79},
respectively,
do not depend on $Y$.
It should be noted that in this NPM the impact of $Y$ on $\chi^2$ is weak,
contrary to what happened in the model of Section~\ref{sec:one},
\textit{cf}.~Fig.~\ref{1multp_OPs_vs_Y}.

\begin{figure}[ht]
\begin{center}
\includegraphics[width=0.5\textwidth]{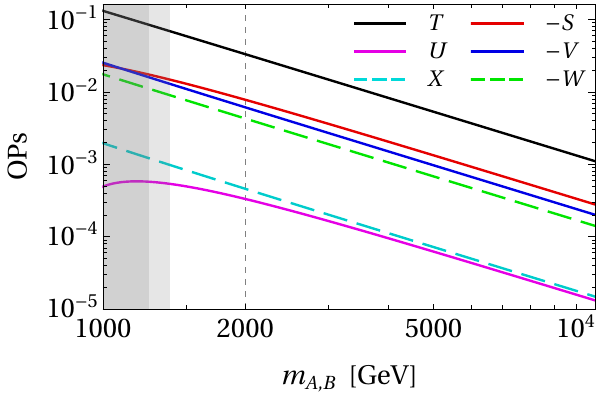}
\end{center}
%\vspace{-6mm}
\caption{The OPs as functions of $m_A = m_B$.
The dashed vertical line indicates the BP2 value $m_{A,B}=2000$\,GeV. 
In the light-gray area,
one obtains $\chi^2 > 17$;
in the dark-gray area,
$\chi^2 > 20$.
For $m_A = m_B \gtrsim 2$\,TeV,
all the OPs vary with $m_A$ as $m_A^{-2}$,
approximately.
%The parameters obey following approximations for $m_{A,B} \gtrsim 2000$\,GeV: $S \approx -21500 \, m_{A,B}^{-1.95}$, $T \approx 132000 \, m_{A,B}^{-2.0}$, $U \approx 770\, m_{A,B}^{-1.92}$, $V \approx -24300\, m_{A,B}^{-2.0}$, $W \approx -17000\, m_{A,B}^{-2.0}$ and $X \approx 1810\, m_{A,B}^{-2.0}$.
}
\label{2multp_Ops_vs_mAB}
\end{figure}

In Fig.~\ref{2multp_Ops_vs_mAB} one observes that,
as the value of $m_A = m_B$ increases,
the absolute values of all the OPs decrease.
Points with very low $m_A = m_B$ tend to have large $\chi^2$;
$\chi^2 \approx 14$ is minimal when $m_A = m_B = 2000$\,GeV
(\ie, at the BP2),
and increases up to $\approx 16$ for larger $m_A = m_B$.
%The darker gray bound indicates
%that Ops give $\chi^2 > 20$ for $m_{A,B} \lesssim 1250$~GeV.
%However, this bound may shift towards smaller masses of $m_{A,B}$
%if the values of $m_C$ and $|m_D|$ are smaller,
%or if the value of $n$ is smaller. 
%For example, when $n=3$,
%the bound of $\chi^2 = 20$ is reached for $m_{A,B} \lesssim 610$~GeV
%with the other BP2 parameters fixed.
%

\begin{figure}[ht]
\begin{center}
\includegraphics[width=1.0\textwidth]{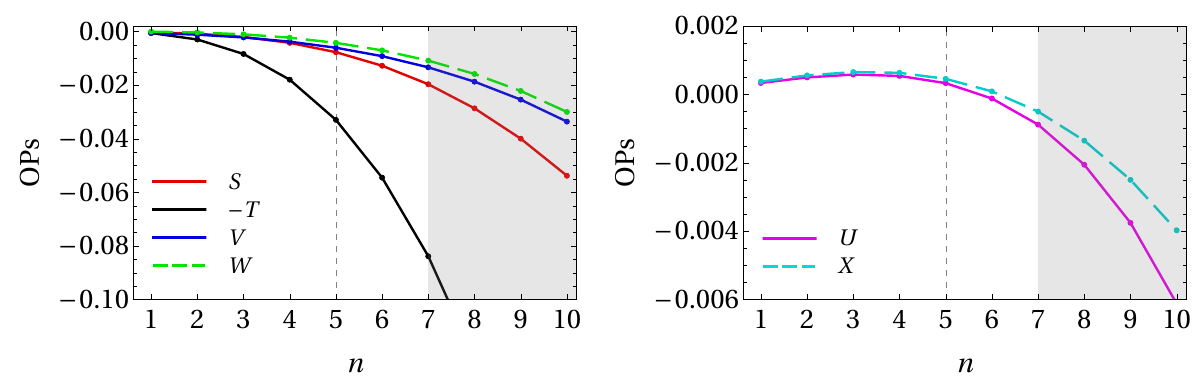}
\end{center}
%\vspace{-6mm}
\caption{The OPs as functions of $n$.
The dashed vertical line indicates the BP2 value $n=5$. 
The gray area corresponds to fits with $\chi^2 > 20$.}
\label{2multp_Ops_vs_n}
\end{figure}

When we keep all the mass parameters and $Y$ fixed at their BP2 values,
and we allow $n$ to vary,
we observe the variation of the OPs displayed in Fig.~\ref{2multp_Ops_vs_n}. 
The absolute values of all the OPs increase with $n$ for $n > 4$,
and eventually $\chi^2$ becomes larger than at the BP2.
%However, a sharper increase is observed
%if the masses of $m_{A,B}$ are smaller,
%or a slower increase if the values of $m_C$ and $|m_D|$ are smaller. 
%

\section{Conclusions}
\label{sec:conclusions}

In this work we have presented general formulas
for all six oblique parameters in an extension of the SM
with additional fermions.
The formulas are based on a formalism which defines matrices $M_L$ and $M_R$
that represent the action of the operator $T_+ \left/ \sqrt{2} \right.$
on the \emph{physical} left- and right-handed fermions,
respectively;
here,
$T_+$ is the raising operator of gauge-$SU(2)$.
Starting from the matrices $M_E$
($E = L, R$)
one calculates the matrices $H_E \equiv \left[ M_E, M_E^\dagger \right]$
and then the matrices $F_E \equiv H_E - 2 Q s_W^2$,
where $Q$ is the electric-charge matrix.
The formulas for the OPs are then Eqs.~\eqref{vw},
\eqref{x},
\eqref{su},
and~\eqref{79},
where one makes use of the functions $\mathcal{F}$,
$\mathcal{G}$,
and $\mathcal{H}$ defined in Eqs.~\eqref{mathcalf},
\eqref{ggg},
and~\eqref{uh},
respectively,
and of functions defined in section~\ref{sec:functions}.

We have applied our formulas to the cases of two models with new
vector-like fermions in arbitrarily large representations of $SU(2)$.
Remarkably,
in both models we have found that the oblique parameters $V$ and $W$
are usually of the same order of magnitude as $S$,
while the oblique parameters $X$ and $U$ tend to be somewhat smaller;
however,
these features may be upended when one is dealing
with fermion representations featuring either a large isospin
$J \gtrsim 2$ or a large hypercharge $\left| Y \right| \gtrsim 5$.

It is worth remarking that,
in the original formulation of the OPs
(see Appendix~\ref{appa}),
the parameters $V$,
$W$,
and $X$ were set to zero
and the parameters $S$ and $U$
had different definitions---$S^\prime$ and $U^\prime$,
respectively.
Our work demonstrates that,
in general,
that original formulation may lead to bad misjudgements,
because neither $V$ and $W$ are necessarily smaller than $S$,
nor necessarily $S \approx S^\prime$ and $U \approx U^\prime$---as is shown
through a simple example in Appendix~\ref{appa}.

\vspace*{5mm}

\paragraph{Acknowledgements:}
L.L.\ thanks Heather Logan for pointing out to him Ref.~\cite{barducci}.
The work of F.A.\ was supported by grant UI/BD/153763/2022.
F.A.\ and L.L.\ were supported by the
Portuguese Foundation for Science and Technology
through projects UIDB/00777/2020,
UIDP/00777/2020,
and CERN/FIS-PAR/0002/2021;
L.L.\ was furthermore supported by CERN/FIS-PAR/0019/2021.
D.J.\ was supported by the Lithuanian Particle Physics Consortium.

\newpage

\begin{appendix}

\setcounter{equation}{0}
\renewcommand{\theequation}{A\arabic{equation}}

\section{The Peskin--Takeuchi approximation}
\label{appa}

In their original formulation by Peskin and Takeuchi~\cite{PT},
the OPs had different definitions.
The parameters $V$,
$W$,
and $X$ did not exist
(or, equivalently, they were set to zero);
the parameter $T$ had the same definition as in Eq.~\eqref{tdef};
and the parameters $S$ and $U$ were instead defined
to be $S^\prime$ and $U^\prime$,
respectively,
where
\bs
\ba
S^\prime &=& \frac{4 s_W^2 c_W^2}{\alpha}
\left[ A^\prime_{ZZ} \left( 0 \right)
  + \frac{c_W^2 - s_W^2}{c_W s_W}\, A^\prime_{\gamma Z} \left( 0 \right)
  - A^\prime_{\gamma \gamma} \left( 0 \right) \right],
\label{sprimedef} \\
U^\prime &=& - S^\prime
+ \frac{4 s_W^2}{\alpha} \left[ A^\prime_{WW} \left( 0 \right)
  + \frac{c_W}{s_W}\, A^\prime_{\gamma Z} \left( 0 \right)
  - A^\prime_{\gamma \gamma} \left( 0 \right) \right].
\label{uprimedef}
\ea
\es
%
%It is clear that,
%if the second derivatives of the functions $A_{ZZ} \left( q^2 \right)$
%and $A_{WW} \left( q^2 \right)$ vanish,
%namely if
%$A^\prime_{ZZ} \left( m_Z^2 \right) \approx A^\prime_{ZZ} \left( 0 \right)$
%and
%$A^\prime_{WW} \left( m_W^2 \right) \approx A^\prime_{WW} \left( 0 \right)$,
%then
%%
%\bs
%\ba
%S^\prime &\approx& S + 4 s_W^2 c_W^2 V,
%\\
%U^\prime &\approx& U + 4 s_W^2 \left( W - c_W^2 V \right).
%\ea
%\es
%%
%Accordingly,
%one may define the asymmetries
%%
%\bs
%\ba
%a_S &\equiv& \frac{S^\prime - S - 4 s_W^2 c_W^2 V}{S^\prime
%  + S + 4 s_W^2 c_W^2 V},
%\\
%a_U &\equiv& \frac{U^\prime - U - 4 s_W^2 \left( W - c_W^2 V \right)}{U^\prime
%  + U + 4 s_W^2 \left( W - c_W^2 V \right)}.
%\ea
%\es
%%
%%%%% I ADDED:
It is clear that
\bs
\ba
S - S^\prime &=& \frac{4 s_W^2 c_W^2}{\alpha} \left[
  \widetilde{A}_{ZZ} \left( m_Z^2 \right) - A^\prime_{ZZ} \left( 0 \right)
  \right],
\\
\left( U + S \right) - \left( U^\prime + S^\prime \right) &=&
\frac{4 s_W^2}{\alpha} \left[
  \widetilde{A}_{WW} \left( m_W^2 \right) - A^\prime_{WW} \left( 0 \right)
  \right].
\ea
\es
%
%%%%%

In our NPM with additional fermions
the Peskin--Takeuchi parameters $S^\prime$ and $U^\prime$ are given by
\bs
\label{supeskin}
\ba
S^\prime \!\!\! &=& \!\!\! \frac{1}{2 \pi}\, \sum_{f, f^\prime}\,
\overline{\mathcal{G}}
\left[ \left( H_L \right)_{f f^\prime}, \left( H_R \right)_{f f^\prime},
  m_f^2, m_{f^\prime}^2 \right]
+ \frac{1}{\pi}\, \sum_f\, Q_f \left( H_L + H_R \right)_{ff}\,
h \left( m_f^2 \right),
\label{ssspeskin}
\\
U^\prime \!\!\! &=& \!\!\! - S^\prime
+ \frac{1}{\pi}\, \sum_{f, f^\prime}\,
\overline{\mathcal{G}}
\left[ \left( M_L \right)_{f f^\prime}, \left( M_R \right)_{f f^\prime},
  m_f^2, m_{f^\prime}^2 \right]
+ \frac{1}{\pi}\, \sum_f\, Q_f \left( H_L + H_R \right)_{ff}\,
h \left( m_f^2 \right),
\hspace*{10mm}
\ea
\es
where
\be
\overline{\mathcal{G}} \left( x, y, I, J \right) = \left( \left| x \right|^2
+ \left| y \right|^2 \right)\, \overline{g} \left( I, J \right)
- 2\, \mathrm{Re} \left( x y^\ast \right)\,
\overline{\hat g} \left( I, J \right) \sqrt{I J},
\label{gggpeskin}
\ee
with functions $\overline{g} \left( I, J \right)$,
$\overline{\hat g} \left( I, J \right)$,
and $h \left( I \right)$ given in Eqs.~\eqref{g000},
\eqref{hatg000},
and~\eqref{6f},
respectively.

For instance,
in the model of section~\ref{sec:one},
wherein $M_L = M_R$,
$H_L = H_R$,
and all the extra fermions have equal mass,
\bs
\ba
S^\prime \!\!\! &=& \!\!\! \frac{1}{\pi}\, \sum_f \left\{
2 Q_f \left( H_L \right)_{ff} h \left( m^2 \right)
+ \left| \left( H_L \right)_{ff} \right|^2
\left[ \overline{g} \left( m^2, m^2 \right)
  - m^2\ \overline{\hat g} \left( m^2, m^2 \right) \right]
\right\},
\\
S^\prime + U^\prime \!\!\! &=& \!\!\! \frac{2}{\pi} \left\{
\sum_f Q_f \left( H_L \right)_{ff} h \left( m^2 \right)
+ \sum_{f, f^\prime} \left| \left( M_L \right)_{f f^\prime} \right|^2
\left[ \overline{g} \left( m^2, m^2 \right)
  - m^2\ \overline{\hat g} \left( m^2, m^2 \right) \right]
\right\}.
\no & &
\ea
\es
Now,
because of Eqs.~\eqref{6f},
\eqref{g000},
and~\eqref{hatg000},
\be
\overline{g} \left( m^2, m^2 \right)
- m^2\ \overline{\hat g} \left( m^2, m^2 \right)
= - h \left( m^2 \right).
\ee
Hence,
\bs
\ba
S^\prime &\propto& \sum_f \left[
2 Q_f \left( H_L \right)_{ff} - \left| \left( H_L \right)_{ff} \right|^2
\right],
\\
S^\prime + U^\prime &\propto& \sum_f Q_f \left( H_L \right)_{ff}
- \sum_{f, f^\prime} \left| \left( M_L \right)_{ff^\prime} \right|^2.
\ea
\es
Using Eqs.~\eqref{j1}--\eqref{j3} one easily concludes
that $S^\prime = U^\prime = 0$ in that NPM.
On the other hand,
in the same NPM the OPs $S$ and $U$ are clearly nonzero---they are not even
necessarily very small.
So,
it is clear that the Peskin--Takeuchi parameters $S^\prime$ and $U^\prime$
do not need to be,
in general,
good approximations to $S$ and $U$,
respectively.

%%%%% I ADDED:
Further dealing on the model of section~\ref{sec:one},
we note that in that model there is only one mass scale,
\textit{viz.}\ the mass $m$ of the new fermions.
Therefore,
since the function $A_{ZZ} \left( q^2 \right)$ has mass-squared dimensions,
\be
A_{ZZ} \left( q^2 \right) = a m^2 + b q^2
+ c\, \frac{q^4}{m^2} + d\, \frac{q^6}{m^4}
+ \mathrm{O} \left( \frac{q^8}{m^6} \right),
\ee
with numerical coefficients $a, b, c, d, \ldots$.
Hence,
\be
\widetilde{A}_{ZZ} \left( m_Z^2 \right) - A^\prime_{ZZ} \left( 0 \right)
= c\, \frac{m_Z^2}{m^2} + d\, \frac{m_Z^4}{m^4}
+ \mathrm{O} \left( \frac{m_Z^6}{m^6} \right).
\ee
This explains the form of Eqs.~\eqref{bsapp}--\eqref{asapp}
in section~\ref{sec:one}.
%%%%%

\end{appendix}

\end{document}